%% file: part_two.tex
\documentclass{article}
\usepackage[T1]{fontenc}
\usepackage[utf8]{inputenc}
\usepackage[style=chem-angew, backend=biber, natbib=true, sorting=none]{biblatex} %
\usepackage[margin=1in]{geometry}
\usepackage[hidelinks]{hyperref}
\usepackage{color}
\usepackage{graphicx}
\usepackage[table,dvipsnames]{xcolor}
\usepackage{enumerate}
\usepackage{setspace}
\usepackage{mhchem}
\usepackage{url}
\usepackage{framed}
\usepackage{todonotes}
\usepackage{placeins}
\usepackage{longtable}
\usepackage{graphbox}
\usepackage{caption}
\usepackage{soul} %
\usepackage{tikz-cd}

\tikzcdset{column sep/normal=1.2em} %

\author{Connor W. Coley\thanks{Department of Chemical Engineering, Massachusetts Institute of Technology, Cambridge, MA 02139}~\footnote{ccoley@mit.edu}, ~ Natalie S. Eyke\footnotemark[1], ~ Klavs F. Jensen\footnotemark[1]~\footnote{kfjensen@mit.edu}}

\date{~}

\newcommand{\challenge}[1]{\noindent \textbf{\textsc{Challenge: } {#1} \addcontentsline{toc}{paragraph}{\it C: #1} \newline \noindent}}

\newcolumntype{L}[1]{>{\raggedright\let\newline\\\arraybackslash\hspace{0pt}}m{#1}}
\newcolumntype{C}[1]{>{\centering\let\newline\\\arraybackslash\hspace{0pt}}m{#1}}
\newcolumntype{R}[1]{>{\raggedleft\let\newline\\\arraybackslash\hspace{0pt}}m{#1}}

\newcommand{\lnameref}[1]{%
\bgroup
\let\nmu\MakeLowercase
\nameref{#1}\egroup}

\DeclareCiteCommand{\citenum}
  {}
  {\printfield{labelnumber}}
  {}
  {}

\title{Autonomous discovery in the chemical sciences part II: Outlook}

\newif\iffull
\fulltrue   

\begin{document}
\maketitle

\input{sections/frontstuff}

\section{Abstract} 

This two-part review examines how automation has contributed to different aspects of discovery in the chemical sciences. In this second part, we reflect on a selection of exemplary studies. It is increasingly important to articulate what the role of automation and computation has been in the scientific process and how that has or has not accelerated discovery. One can argue that even the best automated systems have yet to ``discover'' despite being incredibly useful as laboratory assistants. We must carefully consider how they have been and can be applied to future problems of chemical discovery in order to effectively design and interact with future autonomous platforms.

The majority of this article defines  a large set of open research directions, including improving our ability to work with complex data, build empirical models, automate both physical and computational experiments for validation, select experiments, and evaluate whether we are making progress toward the ultimate goal of autonomous discovery. Addressing these practical and methodological challenges will greatly advance the extent to which autonomous systems can make meaningful discoveries.

\input{sections/trends}

\input{sections/conclusion}

\input{sections/acknowledgements}

\iffull 
\singlespacing
\printbibliography
\fi

\end{document}

%% file: sections/frontstuff.tex
\noindent {\bf Keywords:} automation, chemoinformatics, machine learning, drug discovery, materials science \\[1cm]

\clearpage

\tableofcontents
\doublespacing
\newpage

%% file: sections/trends.tex
\section{Reflection on case studies} %

In 2009, \citeauthor{king_automation_2009} proposed a hypothetical, independent robot scientist that "automatically originates hypotheses to explain observations, devises experiments to test these hypotheses, physically runs the experiments by using laboratory robotics, interprets the results, and then repeats the cycle" \cite{king_automation_2009}. To what extent have we closed the gap toward each of these workflow components, and what challenges remain? 

The case studies in Part 1 illustrate many examples of the progress that has been made toward achieving machine autonomy in discovery. Several studies in particular, summarized in \autoref{tab:discovery_examples}, represent what we consider to be exemplars of different discovery paradigms \cite{williams_cheaper_2015, king_functional_2004,weber_optimization_1995, desai_rapid_2013, j.reizman_suzukimiyaura_2016, nikolaev_discovery_2014, kangas_efficient_2014, gao_reaction_2016, fang_discovery_2016, gomez-bombarelli_design_2016, Segler2018, thornton_materials_2017, janet_accelerating_2018, macleod_self-driving_2019}. These include the successful execution of experimental and computational workflows as well as the successful implementation of automated experimental selection and belief revision.  There are a great number of studies (some of which are described in part one) that follow the paradigm of (a) train a surrogate QSAR/QSPR model on existing data, (b) computationally design a molecule or material to optimize predicted performance, and (c) manually validate a few compounds. The table intentionally underrepresents such studies, as we believe iterative validation to be a distinguishing feature of autonomous workflows compared to ``merely'' automated calculations.

\begin{table}[h]
    \centering
    
    \scriptsize
    \begin{tikzcd}[column sep=4em, row sep=3em]
    \text{initial design}
    \arrow[black,bend left=20]{r}[black,above]{\text{algorithmic}}
    \arrow[black,bend right=20]{r}[black,below]{\text{expert-defined}}
     & 
    \text{selected experiment} 
    \arrow[black,bend left=20]{r}[black,above]{\text{physical}}
    \arrow[black,bend right=20]{r}[black,below]{\text{computational}}
    \arrow[black,bend right=70,<-]{r}[black,below]{\text{model-free design}}
    \arrow[black,bend left=40,<-]{rr}[black,above]{\text{model-based design}}
    &
    \text{data}
    \arrow[black]{r}[black,above]{\text{update}}
    &
    \text{belief/model}
    \arrow[black,<-]{r}[black,above]{\text{initialize}}
    &
    \text{existing data}
    \end{tikzcd}\vspace{0.5cm}

    \tiny
    \rowcolors{2}{Apricot!5}{Apricot!25}
    \begin{tabular}{L{1.2cm}|L{1.75cm}|L{2cm}|L{2cm}|L{2cm}|L{3cm}|l}
        \hline 
         Reference & Discovery & Initialization & Design space & Data {generation} & Notes & Workflow \\ \hline

         Eve \cite{williams_cheaper_2015} & 
         bioactive, selective molecules & 4,800 random compounds from design space & 
         fixed library of 14,400 compounds & automated measurement of yeast growth curves & 
         compound screening from a fixed library with an active search &
        \begin{tikzcd} \cdot
        \arrow[red,bend left=20]{r}{} %
        \arrow[black,bend right=20]{r}{} %
        &[0.5em] \cdot
        \arrow[red,bend left=20]{r}{} %
        \arrow[black,bend right=20]{r}{} %
        \arrow[black,bend right=70,<-]{r}{} %
        \arrow[red,bend left=40,<-]{rr}{} %
        &[0.5em] \cdot
        \arrow[red]{r}{} %
        &[0.25em] \cdot
        \arrow[black,<-]{r}{} %
        &[0.25em] \cdot \end{tikzcd} \\

         Adam \cite{king_functional_2004}  & 
         gene-enzyme relationships & random experiment & 
         15 open reading frame deletions & automated auxotrophy experiments & 
         narrow hypothesis space, but nearly-closed-loop experimentation & 
        \begin{tikzcd} \cdot
        \arrow[red,bend left=20]{r}{} %
        \arrow[black,bend right=20]{r}{} %
        &[0.5em] \cdot
        \arrow[red,bend left=20]{r}{} %
        \arrow[black,bend right=20]{r}{} %
        \arrow[black,bend right=70,<-]{r}{} %
        \arrow[red,bend left=40,<-]{rr}{} %
        &[0.5em] \cdot
        \arrow[red]{r}{} %
        &[0.25em] \cdot
        \arrow[black,<-]{r}{} %
        &[0.25em] \cdot \end{tikzcd} \\

         \citeauthor{weber_optimization_1995} \cite{weber_optimization_1995} & 
         thrombin inhibitors & 20 random compounds & 
         virtual $10 \times 10 \times 40 \times 40$ compound library &
         manual synthesis and inhibition assay &
         iterative optimization using a genetic algorithm; design space defined by 4-component Ugi reaction to ensure synthesizability & 
        \begin{tikzcd} \cdot
        \arrow[red,bend left=20]{r}{} %
        \arrow[black,bend right=20]{r}{} %
        &[0.5em] \cdot
        \arrow[red,bend left=20]{r}{} %
        \arrow[black,bend right=20]{r}{} %
        \arrow[red,bend right=70,<-]{r}{} %
        \arrow[black,bend left=40,<-]{rr}{} %
        &[0.5em] \cdot
        \arrow[black]{r}{} %
        &[0.25em] \cdot
        \arrow[black,<-]{r}{} %
        &[0.25em] \cdot \end{tikzcd} \\
         
         \citeauthor{desai_rapid_2013} \cite{desai_rapid_2013} & 
         kinase inhibitors & random compound & 
         $27 \times 10$ candidates in make-on-demand library & 
         automated microfluidic synthesis and biological testing & 
         closed-loop synthesis and biological testing; narrow chemical design space &
        \begin{tikzcd} \cdot
        \arrow[red,bend left=20]{r}{} %
        \arrow[black,bend right=20]{r}{} %
        &[0.5em] \cdot
        \arrow[red,bend left=20]{r}{} %
        \arrow[black,bend right=20]{r}{} %
        \arrow[black,bend right=70,<-]{r}{} %
        \arrow[red,bend left=40,<-]{rr}{} %
        &[0.5em] \cdot
        \arrow[red]{r}{} %
        &[0.25em] \cdot
        \arrow[black,<-]{r}{} %
        &[0.25em] \cdot \end{tikzcd} \\
         
          \citeauthor{j.reizman_suzukimiyaura_2016} \cite{j.reizman_suzukimiyaura_2016} & reaction conditions & algorithmic D-optimal design & concentration, temperature, time, 8 catalysts & automated synthesis and yield quantitation & closed-loop reaction optimization through a screening phase and an iterative phase  & 
        \begin{tikzcd} \cdot
        \arrow[red,bend left=20]{r}{} %
        \arrow[black,bend right=20]{r}{} %
        &[0.5em] \cdot
        \arrow[red,bend left=20]{r}{} %
        \arrow[black,bend right=20]{r}{} %
        \arrow[black,bend right=70,<-]{r}{} %
        \arrow[red,bend left=40,<-]{rr}{} %
        &[0.5em] \cdot
        \arrow[red]{r}{} %
        &[0.25em] \cdot
        \arrow[black,<-]{r}{} %
        &[0.25em] \cdot \end{tikzcd} \\

         ARES \cite{nikolaev_discovery_2014, nikolaev_autonomy_2016} & 
         carbon nanotube growth conditions & 
         84 expert-defined experimental conditions & 
         process conditions (temperature, pressures, and gas compositions) & automated nanotube growth and characterization & complex experimentation; uses RF model for regression & 
        \begin{tikzcd} \cdot
        \arrow[black,bend left=20]{r}{} %
        \arrow[red,bend right=20]{r}{} %
        &[0.5em] \cdot
        \arrow[red,bend left=20]{r}{} %
        \arrow[black,bend right=20]{r}{} %
        \arrow[black,bend right=70,<-]{r}{} %
        \arrow[red,bend left=40,<-]{rr}{} %
        &[0.5em] \cdot
        \arrow[red]{r}{} %
        &[0.25em] \cdot
        \arrow[black,<-]{r}{} %
        &[0.25em] \cdot \end{tikzcd} \\

        \citeauthor{kangas_efficient_2014} \cite{kangas_efficient_2014} & bioactive compounds against many assays &
        384 random measurements from design space & 177 assay $\times$ 20,000 compound interactions & 
        \emph{simulated} experiments by revealing PubChem measurements &
        validation of pool-based active learning framework through retrospective analysis; iterative batches of 384 experiments &
        \begin{tikzcd} \cdot
        \arrow[red,bend left=20]{r}{} %
        \arrow[black,bend right=20]{r}{} %
        &[0.5em] \cdot
        \arrow[red,bend left=20]{r}{} %
        \arrow[black,bend right=20]{r}{} %
        \arrow[black,bend right=70,<-]{r}{} %
        \arrow[red,bend left=40,<-]{rr}{} %
        &[0.5em] \cdot
        \arrow[red]{r}{} %
        &[0.25em] \cdot
        \arrow[black,<-]{r}{} %
        &[0.25em] \cdot \end{tikzcd} \\

         RMG  \cite{gao_reaction_2016} & detailed gas-phase kinetic mechanisms &  reaction conditions, optionally seeded by known mechanism & elementary reactions following expert-defined reaction templates & estimation of thermodynamic and kinetic parameters & iterative addition of hypothesized elementary reactions to a kinetic model based on simulations using intermediate models & 
        \begin{tikzcd} \cdot
        \arrow[red,bend left=20]{r}{} %
        \arrow[black,bend right=20]{r}{} %
        &[0.5em] \cdot
        \arrow[black,bend left=20]{r}{} %
        \arrow[red,bend right=20]{r}{} %
        \arrow[black,bend right=70,<-]{r}{} %
        \arrow[red,bend left=40,<-]{rr}{} %
        &[0.5em] \cdot
        \arrow[red]{r}{} %
        &[0.25em] \cdot
        \arrow[red,<-]{r}{} %
        &[0.25em] \cdot \end{tikzcd} \\
          
          \citeauthor{fang_discovery_2016} \cite{fang_discovery_2016} & neuroprotective compounds & activity data from ChEMBL & in-house library of 28k candidates & none & literature-trained QSAR model applied to noniterative virtual screening with manual \emph{in vitro} validation & 
        \begin{tikzcd} \cdot
        \arrow[black,bend left=20]{r}{} %
        \arrow[black,bend right=20]{r}{} %
        &[0.5em] \cdot
        \arrow[red,bend left=20]{r}{} %
        \arrow[black,bend right=20]{r}{} %
        \arrow[black,bend right=70,<-]{r}{} %
        \arrow[red,bend left=40,<-]{rr}{} %
        &[0.5em] \cdot
        \arrow[black]{r}{} %
        &[0.25em] \cdot
        \arrow[red,<-]{r}{} %
        &[0.25em] \cdot \end{tikzcd} \\

         \citeauthor{gomez-bombarelli_design_2016} \cite{gomez-bombarelli_design_2016} & organic light-emitting diode molecules &  40k random compounds from design space & virtual library of 1.6 M enumerated compounds & DFT calculations & iteratively selected batches of 40k calculations; manually validated a small number of compounds experimentally & 
        \begin{tikzcd} \cdot
        \arrow[red,bend left=20]{r}{} %
        \arrow[black,bend right=20]{r}{} %
        &[0.5em] \cdot
        \arrow[black,bend left=20]{r}{} %
        \arrow[red,bend right=20]{r}{} %
        \arrow[black,bend right=70,<-]{r}{} %
        \arrow[red,bend left=40,<-]{rr}{} %
        &[0.5em] \cdot
        \arrow[red]{r}{} %
        &[0.25em] \cdot
        \arrow[black,<-]{r}{} %
        &[0.25em] \cdot \end{tikzcd} \\

        \citeauthor{Segler2018} \cite{Segler2018} & bioactive compounds &  1.4M molecules from ChEMBL & all of chemical space & surrogate QSAR/QSPR models of activity & iteratively refined generative LSTM model (pretrained on ChEMBL) on active molecules identified via sampling an initial 100k + 8 rounds $\times$ 10k molecules & 
        \begin{tikzcd} \cdot
        \arrow[black,bend left=20]{r}{} %
        \arrow[black,bend right=20]{r}{} %
        &[0.5em] \cdot
        \arrow[black,bend left=20]{r}{} %
        \arrow[red,bend right=20]{r}{} %
        \arrow[black,bend right=70,<-]{r}{} %
        \arrow[red,bend left=40,<-]{rr}{} %
        &[0.5em] \cdot
        \arrow[red]{r}{} %
        &[0.25em] \cdot
        \arrow[red,<-]{r}{} %
        &[0.25em] \cdot \end{tikzcd} \\

         \citeauthor{thornton_materials_2017} \cite{thornton_materials_2017} & hydrogen storage materials & 
         200 human-chosen subset and 200 random subset of search space & 
         850k structures (Materials Genome) & 
         grand canonical Monte Carlo simulations &
         few rounds of greedy optimization with batches of 1000 using surrogate QSPR model  &
        \begin{tikzcd} \cdot
        \arrow[black,bend left=20]{r}{} %
        \arrow[red,bend right=20]{r}{} %
        &[0.5em] \cdot
        \arrow[black,bend left=20]{r}{} %
        \arrow[red,bend right=20]{r}{} %
        \arrow[black,bend right=70,<-]{r}{} %
        \arrow[red,bend left=40,<-]{rr}{} %
        &[0.5em] \cdot
        \arrow[red]{r}{} %
        &[0.25em] \cdot
        \arrow[black,<-]{r}{} %
        &[0.25em] \cdot \end{tikzcd} \\
         
          \citeauthor{janet_accelerating_2018} \cite{janet_accelerating_2018} & spin-state splitting inorganic complexes & random complexes from design space & 708 ligand combinations $\times$ 8 transition metals & ANN surrogate model prediction & used genetic algorithm and computational evaluation for iterative optimization; relies on ANN pretrained on 2690 DFT calculations &
        \begin{tikzcd} \cdot
        \arrow[red,bend left=20]{r}{} %
        \arrow[black,bend right=20]{r}{} %
        &[0.5em] \cdot
        \arrow[black,bend left=20]{r}{} %
        \arrow[red,bend right=20]{r}{} %
        \arrow[red,bend right=70,<-]{r}{} %
        \arrow[black,bend left=40,<-]{rr}{} %
        &[0.5em] \cdot
        \arrow[black]{r}{} %
        &[0.25em] \cdot
        \arrow[black,<-]{r}{} %
        &[0.25em] \cdot \end{tikzcd} \\

          Ada \cite{macleod_self-driving_2019} & organic hole transport materials & algorithmic & dopant ratio and annealing time & automated synthesis and analysis of thin film & complex experimentation successfully automated; simple design space &
        \begin{tikzcd} \cdot
        \arrow[red,bend left=20]{r}{} %
        \arrow[black,bend right=20]{r}{} %
        &[0.5em] \cdot
        \arrow[red,bend left=20]{r}{} %
        \arrow[black,bend right=20]{r}{} %
        \arrow[black,bend right=70,<-]{r}{} %
        \arrow[red,bend left=40,<-]{rr}{} %
        &[0.5em] \cdot
        \arrow[red]{r}{} %
        &[0.25em] \cdot
        \arrow[black,<-]{r}{} %
        &[0.25em] \cdot \end{tikzcd} \\
         
         \hline 

    \end{tabular}
    \caption{Selected examples of discovery accelerated by automation or computer assistance. The stages of the discovery workflow employed by each are shown as red arrows corresponding to the schematic above. Workflows may begin either with an initial set of experiments to run or by initializing a model with existing (external) data. Algorithmic initial designs include the selection of random experiments from the design space.} %
    \label{tab:discovery_examples}
\end{table}

We encourage the reader to reflect on these case studies through the lens of the questions we proposed for assessing autonomous discovery: (i) How broadly is the goal defined? (ii) How constrained is the search/design space? (iii) How are experiments for validation/feedback selected? (iv) How superior to a brute force search is navigation of the design space? (v) How are experiments for validation/feedback performed? (vi) How are results organized and interpreted? (vii) Does the discovery outcome contribute to broader scientific knowledge?

The goals of discovery are defined narrowly in most studies. We are not able to request that a platform identify a good therapeutic, come up with an interesting material, uncover a new reaction, or propose an interesting model. Instead, in most studies described to date, an expert defines a specific scalar performance objective that an algorithm tries to optimize. Out of the examples in Table~\ref{tab:discovery_examples}, \citeauthor{kangas_efficient_2014} has the highest-level goals: in one of their active learning evaluations, the goal could be described as finding strong activity for \emph{any} of the 20,000 compounds against \emph{any} of the 177 assays. 
While Adam attempts to find relationships between genes and the enzymes they encode (discover a causal model), it does so from a very small pool of hypotheses for the sake of compatibility with existing deletion mutants for a single yeast strain. 

The search spaces used in these studies vary widely in terms of the constraints that are imposed upon them. Some are restricted out of necessity to ensure validation is automatable (e.g., Eve, Adam, \citeauthor{desai_rapid_2013}, ARES) or convenient (e.g., \citeauthor{weber_optimization_1995}, \citeauthor{fang_discovery_2016}). Others constrain the search space to a greater degree than  automated validation requires. This includes reductions of 1) dimensionality, by holding process parameters constant (e.g., \citeauthor{j.reizman_suzukimiyaura_2016}, Ada), or 2) the size of discrete candidate spaces (e.g., \citeauthor{gomez-bombarelli_design_2016}, \citeauthor{thornton_materials_2017}, \citeauthor{janet_accelerating_2018}). Computational studies that minimize constraints on their search (e.g., RMG, \citeauthor{Segler2018}) do so under the assumption that the results of validation (e.g., simulation results, predictions from surrogate models) will be accurate across the full design space.

In all cases, human operators have implicitly or explicitly assumed that a good solution can be found in these restricted spaces. The extent to which domain expertise or prior knowledge is needed to establish the design space also varies. Molecular or materials design in relatively unbounded search spaces (e.g., \citeauthor{Segler2018}) requires the least human input.  Fixed candidate libraries that are small (e.g., \citeauthor{weber_optimization_1995}, \citeauthor{desai_rapid_2013}) or derived from an expert-defined focused enumeration (e.g., RMG, \citeauthor{gomez-bombarelli_design_2016}, \citeauthor{janet_accelerating_2018}) require significant application-specific domain knowledge; larger fixed candidate libraries may be application-agnostic (e.g., diverse screening libraries in Eve, \citeauthor{fang_discovery_2016}, \citeauthor{thornton_materials_2017}). Limiting process parameters (e.g., \citeauthor{j.reizman_suzukimiyaura_2016}, ARES, Ada) require more general knowledge about typical parameter ranges where optima may lie.

The third question regarding experiment selection is one where the field has excelled. There are many frameworks for quantifying the value of an experiment in model-guided experimental design, both when optimizing for performance and when optimizing for information \cite{settles2012active}. However, active learning with formal consideration of uncertainty from either a frequentist perspective \cite{anderson-cook_response_2009} (e.g., Eve, \citeauthor{j.reizman_suzukimiyaura_2016}) or a Bayesian perspective \cite{Mockus1978,frazier2018tutorial} (e.g., Ada) is less common than with \emph{ad hoc} definitions of experimental diversity  meant to encourage exploration (e.g., \citeauthor{desai_rapid_2013}, \citeauthor{kangas_efficient_2014}). Both are less common than greedy selection criteria (e.g., ARES, RMG, \citeauthor{gomez-bombarelli_design_2016}, \citeauthor{thornton_materials_2017}). Model-free experiment selection, including the use of genetic algorithms (e.g., \citeauthor{weber_optimization_1995}, \citeauthor{janet_accelerating_2018}), is also quite prevalent but requires some additional overhead from domain experts to determine allowable mutations within the design space. When validation is not automated (e.g., \citeauthor{fang_discovery_2016} or \citeauthor{gomez-bombarelli_design_2016}'s experimental follow-up), the selection of experiments is best described as pseudo-greedy, where the top predicted candidates are manually evaluated for practical factors like synthesizability.

The magnitude of the benefit of computer-assisted experiment selection is a function of the size of the design space and, when applicable, the initialization required prior to the start of the iterative phase. In many cases, a brute force exploration of the design space is not prohibitively expensive (e.g., Eve, Adam, \citeauthor{desai_rapid_2013}, \citeauthor{janet_accelerating_2018}), although this is harder to quantify when some of the design variables are continuous (e.g., \citeauthor{j.reizman_suzukimiyaura_2016}, ARES, Ada). Other design spaces are infinite or virtually infinite (e.g., RMG, \citeauthor{Segler2018}), which makes the notion of a brute force search ill-defined. Regardless of whether the full design space can be screened, we can still achieve a reduction in the number of experiments required to find high-performing candidates, perhaps by a modest factor of 2-10 (e.g., Eve, \citeauthor{desai_rapid_2013}, \citeauthor{gomez-bombarelli_design_2016}, \citeauthor{janet_accelerating_2018}) or even by 2-3 orders of magnitude (e.g., \citeauthor{weber_optimization_1995}, \citeauthor{kangas_efficient_2014}, \citeauthor{thornton_materials_2017}).  It's possible that the experimental efficiency of some of these studies could be improved by reducing the number of experiments needed to initialize the workflow (e.g., Eve, \citeauthor{gomez-bombarelli_design_2016}).%

The manner in which experiments for validation are performed depends on the nature of the design space and, of course, whether experiments are physical or computational. Examples where validation is automated are intentionally overrepresented in this review; there are \emph{many} more examples of partially-autonomous discovery in which models prioritize experiments that are manually performed (e.g., similar to \citeauthor{weber_optimization_1995}, \citeauthor{fang_discovery_2016}). There are also cases where \emph{almost} all aspects of experimentation are automated but a few manual operations remain (e.g., transferring well plates for Adam). In computational workflows, one can often construct pipelines to automate calculations (e.g., RMG, \citeauthor{gomez-bombarelli_design_2016}, \citeauthor{Segler2018},  \citeauthor{thornton_materials_2017}, \citeauthor{janet_accelerating_2018}). In experimental workflows, one can programmatically set process conditions with tailor-made platforms (e.g., \citeauthor{desai_rapid_2013}, ARES, Ada) or  robotically perform assays using in-stock compounds (e.g., Eve, Adam) or ones synthesized on-demand (e.g., \citeauthor{desai_rapid_2013}). Pool-based active learning strategies lend themselves to retrospective validation, where an ``experiment'' simply reveals a previous result not known to the algorithm (e.g., \citeauthor{kangas_efficient_2014}); this is trivially automated and thus attractive for method development. Note that in the workflow schematic for \citeauthor{kangas_efficient_2014} in Table~\ref{tab:discovery_examples}, we illustrate the revelation of PubChem measurements as experiments.

In iterative workflows, automating the organization and interpretation of results is a practical step toward automating the subsequent selection of experiments. When workflows only proceed through a few iterations of batched experiments, humans may remain in the loop to simplify the logistics of organizing results and initializing each round (e.g., \citeauthor{gomez-bombarelli_design_2016}, \citeauthor{thornton_materials_2017}), but nothing fundamentally prevents this step from being automated. When many iterations are required or expected, it behooves us to ensure that the results of experiments can be directly interpreted; otherwise, tens (e.g., \citeauthor{desai_rapid_2013}, \citeauthor{j.reizman_suzukimiyaura_2016}, \citeauthor{Segler2018}, \citeauthor{janet_accelerating_2018}, Ada) or hundreds (e.g., Eve, ARES, \citeauthor{kangas_efficient_2014}) of interventions by human operators would be required. This can be the case when iterative experimental design is used with manual experimentation and analysis (e.g., the 20 iterations of a genetic algorithm  conducted manually by \citeauthor{weber_optimization_1995}). In non-iterative workflows with manual validation (e.g., \citeauthor{fang_discovery_2016}), there is little benefit to automating the interpretation of new data. Relative to automating experiments and data acquisition, automating the interpretation thereof is rarely an obstacle to autonomy. Exceptions to this include cases where novel physical matter (a molecule or material) is synthesized and characterized (e.g., case studies in Part 1 related to experimental reaction discovery), where further advances in computer-aided structural elucidation (CASE) \cite{engel_structure-spectrum_2018} are needed. 

Our final question when assessing autonomy is whether the discovery outcome contributes to broader scientific knowledge. In Table~\ref{tab:discovery_examples}, with the exception of Adam and RMG, we have focused on the discovery of physical matter or processes rather than models. The primary outcome of  these discovery campaigns is the identification of a molecule, material, or set of process conditions that achieves or optimizes a human-defined performance objective. Workflows with model-based experimental designs (e.g., all but \citeauthor{weber_optimization_1995} and \citeauthor{janet_accelerating_2018}, who use genetic algorithms) have the secondary outcome of a surrogate model, which may or may not lend itself to interpretation. However, the point of this question is whether the contribution to broader scientific knowledge came \emph{directly} from the autonomous platform, not through an \emph{ex post facto} analysis by domain experts. These discoveries generally require manual interpretation, again excepting Adam, RMG, and similar platforms where what is discovered is part of a causal model.

Our first and last questions represent lofty goals in autonomous discovery: we specify high-level goals and receive human-interpretable, generalized conclusions \emph{beyond} the identification of a molecule, material, device, process, or black box model. However, we have made tremendous progress in offloading both the manual and mental burden of navigating design spaces through computational experimental design and automated validation. We often impose constraints on design spaces to avoid unproductive exploration and focus the search on what we believe to be plausible candidates. To widen a design space requires that experiments remain automatable--less of a challenge for computational experiments than for physical ones--but may decrease the need for subject matter expertise and may increase the odds that the platform identifies an unexpected or superior result. Well-established frameworks of active learning and Bayesian optimization have served us well for experimental selection, while new techniques in deep generative modeling have opened up  opportunities for exploring virtually-infinite design spaces.

\FloatBarrier 

\section{Challenges and trends}

The capabilities required for autonomous discovery are coming together rapidly. 
This section emphasizes what we see as key remaining challenges associated with working with complex data, automating validation and feedback, selecting experiments, and evaluation. %

\subsection{Working with complex data}

The discovery of complex phenomena requires a tight connection between knowledge and data \cite{gil_discovery_nodate}. A 1991 article laments the ``growing gap between data generation and data understanding'' and the great potential for knowledge discovery from databases \cite{Piateski:1991:KDD:583310}. While we continue to generate new data at an increasing rate, we have also dramatically improved our ability to make sense of complex datasets through new algorithms and advances in computing power.

We intentionally use ``complex data'' rather than ``big data''--the latter generally refers only to the size or volume of data, and not its content. Here, we mean ``complex data'' when it would be difficult or impossible for a human to identify the same relationships or conclusions as an algorithm. This may be due to the size of the dataset (e.g., millions of bioactivity measurements), the lack of structure (e.g., journal articles), or the dimensionality (e.g., a regression of multidimensional process parameters).

Complex datasets come in many forms and have inspired an array of different algorithms for making sense of them (and leveraging them for discovery). Unstructured data can be mined and converted into structured data \cite{kim_machine-learned_2017, gyori_word_2017} or directly analyzed as text, e.g. to develop hypotheses about new functional materials \cite{tshitoyan_unsupervised_2019}. Empirical models can be generated and used to draw inferences about factors that influence complex chemical reactivity \cite{raccuglia_machine-learning-assisted_2016, DerekTAhneman2018, zahrt_prediction_2019, reid_holistic_2019}. Virtually \emph{any} dataset of (input, output) pairs describing a performance metric of a molecule, material, or process can serve as the basis for supervised learning of a surrogate model. Likewise, unsupervised techniques can be used to infer the structure of complex datasets \cite{f.reinhart_machine_2017,mardt_vampnets_2018, rives_biological_2019} and form the basis of deep generative models that propose new physical matter \cite{elton_deep_2019, schwalbe-koda_generative_2019}.

\subsubsection{Creating and maintaining datasets}

Many studies don't develop substantially novel methods, but instead take advantage of new data resources. This is facilitated by the increasing availability of public databases. 
The PubChem database, maintained by the NIH and currently the largest repository of open-access chemical information \cite{kim_pubchem_2019}, has been leveraged by many studies for ligand-based drug discovery proofs and thus is a particularly noteworthy example of the value inherent in these curation efforts. Curation efforts spearheaded by government organizations as well as those led by individual research groups can both be enormously impactful, whether through amalgamation of large existing datasets (a greater strength of the broad collaborations) or the  accumulation of high-quality, well-curated data (which efforts by individual research groups tend may be better suited for).

\autoref{tab:databases} provides a list of some popular databases used for tasks related to chemical discovery. Additional databases related to materials science can be found in refs.~\citenum{hill_materials_2016} and \citenum{himanen_data-driven_2019}. Some related to drug discovery are contained in refs.~\citenum{rigden_2018_2018} and \citenum{leejunhyun_databases_2019}. Additional publications compare commercial screening libraries that can be useful in experimental or computational workflows \cite{krier_assessing_2006, langdon_scaffold_2011}.

\begin{center}
    \footnotesize  \renewcommand\arraystretch{1.5} \setstretch{0.75}
    \begin{longtable}{p{3cm}|p{9cm}|p{2cm}|p{2cm}}
    \caption{Overview of some databases used to facilitate discovery in the chemical sciences. API: application programming interface}
    \label{tab:databases}\\
         \multicolumn{4}{c}{ Chemical structures  } \\ \hline
         Name &   Description &  Size (approx.) & Availability \\  \hline 
         ZINC \cite{irwin_zinc_2005,sterling_zinc_2015}    &
         commercially-available compounds &
         35 M &
         Open \\
         
         ChemSpider \cite{noauthor_chemspider_nodate}  &
         structures and misc. data &
         67 M &
         API \\
         
         SureChEMBL \cite{papadatos_surechembl:_2016}  &
         structures from ChEMBL &
         1.9 M &
         Open \\
         
         Super Natural II \cite{banerjee_super_2015} & 
         natural product structures &
         325 k & 
         Open \\
         
         SAVI \cite{noauthor_synthetically_nodate}     &
         enumerated synthetically-accessible structures and their building blocks &
         283 M &
         Open \\
         
        eMolecules \cite{emolcules} &
         commercially-available chemicals and prices &
         5.9 M &
         Commercial \\
         
        MolPort \cite{molport} &
        in-stock chemicals and prices &         
        7 M &
        On Request \\
         
        REAL (Enamine) \cite{enamine_REAL} &
         enumerated synthetically-accessible structures &
         11 B &
         Open \\
         
         Chemspace \cite{chemspace} & 
         in-stock chemicals and prices &
         1 M &
         On Request \\

        GDB-11, GDB-13, GDB-17 \cite{fink_virtual_2007, blum_970_2009, ruddigkeit_enumeration_2012}     &
         exhaustively enumerated chemical structures  &
         26.4 M; 970 M; 166 B &
         Open \\

        SCUBIDOO \cite{chevillard_scubidoo:_2015}     &
         enumerated synthetically-accessible structures &
         $>10$ M &
         Open \\
         
        CHIPMUNK \cite{humbeck_chipmunk:_2018}     &
         enumerated synthetically-accessible structures &
         95 M &
         Open \\ \hline

         \multicolumn{4}{c}{ Biological data } \\ \hline
         Name &   Description &  Size & Availability \\  \hline 
         
         PubChem \cite{kim_pubchem_2019}  &
         compounds and properties, emphasis on bioassay results &
         96 M &
         Open \\
         
         ChEMBL \cite{noauthor_chembl_nodate, gaulton_chembl:_2012}  &
         compounds and bioactivity measurements &
         1.9 M  &
         Open \\
         
         ChEBI \cite{hastings_chebi_2016} &
         compounds and biological relevance &
         56 k &
         Open \\
         
         PDB \cite{noauthor_rcsb_nodate}      &
         biological macromolecular structures &
         150 k &
         Open \\
         
         PDBBind \cite{noauthor_welcome_nodate}   &
         protein binding affinity &
         20 k &
         Open \\
         
         ProTherm \cite{gromiha_protherm_2002}      &
         thermodynamic data for proteins &
         10 k &
         Open \\
         
         LINCS \cite{keenan_library_2018}      &
         cellular interactions and perturbations &
         varies &
         Open \\

         SKEMPI \cite{jankauskaite_skempi_2019}      &
         energetics of mutant protein interactions &
         7 k &
         Open \\
         
         xMoDEL \cite{noauthor_xmodel:_nodate}      &
         MD trajectories of proteins &
         1700 &
         Open \\
         
         GenBank \cite{benson_genbank_2018}      &
         species' nucleotide sequences &
         400 k &
         Open \\
         
         DrugBank \cite{Wishart2006} &
         drug compounds, associated chemical properties, and pharmacological information &
         13 k &
         Open \\
         
         BindingDB \cite{gilson_bindingdb_2016} & 
         compounds and binding measurements & 
         750 k; 1.7 M &
         Open \\
         
         CDD  \cite{ekins2013collaborative} & %
         collaborative drug discovery database for neglected tropical diseases  &
         $>100$ datasets &
         Registration \\

         ToxCast \cite{dix2007toxcast, richard_toxcast_2016} &
         compounds and cellular responses & 
         $>4500$ & 
         Open \\

         Tox21 \cite{tice2013improving, tox21} &
         compounds and multiple bioassays  & 
         14k & 
         Open \\ \hline 
         \multicolumn{4}{c}{ Chemical reactions  } \\ \hline
         Name &   Description &  Size & Availability \\  \hline 
         USPTO \cite{lowe_chemical_2017}    &
         chemical reactions (patent literature) &
         3.3 M &
         Open \\
         
         Pistachio \cite{pistachio} &
         chemical reactions (patent literature) &
         8.4 M &
         Commercial \\
         
         Reaxys \cite{noauthor_reaxys_nodate}    &
         chemical reactions &
         $>$10 M &
         Commercial \\
         
         CASREACT \cite{noauthor_reactions_nodate}   &
         chemical reactions &
         $>$10 M &
         Commercial \\
         
         SPRESI \cite{noauthor_infochem_nodate}    &
         chemical reactions &
         4.3 M &
         Commercial \\
         
         Organic Reactions  \cite{noauthor_databases_nodate}    & %
         chemical reactions &
         250 k &
         Commercial \\
         
         EAWAG-BBD \cite{gao_university_2010}   &
         biocatalysis and biodegradation pathways & %
         219 & %
         Open \\
         
         NIST Chemical Kinetics \cite{nist_kinetics} & 
         gas-phase chemical reactions & 
         38 k &
         Open \\

        NMRShiftDB \cite{steinbeck_nmrshiftdb_2004} & 
        measured NMR spectra & 
        52 k &
        Open \\ \hline 

         \multicolumn{4}{c}{ Molecular properties } \\ \hline
         Name &   Description &  Size & Availability \\  \hline 
         QM7/QM7b \cite{rupp_fast_2012,montavon_machine_2013}  &
         electronic properties (DFT) &
         7200 &
         Open \\
         
         QM9 \cite{ramakrishnan_quantum_2014}  &
         electronic properties (DFT) &
         134 k &
         Open \\
         
         QM8 \cite{ramakrishnan_electronic_2015}  &
         spectra and excited state properties &
         22 k &
         Open \\

         FreeSolv \cite{mobley_freesolv:_2014} &
         aqueous solvatio energies &
         642 & 
         Open \\
         
         NIST Chemistry WebBook \cite{linstrom_nist_2001} & 
         miscellaneous molecular properties & 
         varies &
         Open   \\ \hline

         \multicolumn{4}{c}{ Materials } \\ \hline
         Name &   Description &  Size & Availability \\  \hline 

         PoLyInfo \cite{otsuka2011polyinfo}  &
         polymer properties &
         400k &
         Open \\

         COD \cite{Merkys2016,Grazulis2015,Grazulis2012}  &
         crystal structures of organic, inorganic, metal-organics compounds and minerals &
         410k &
         Open \\
         
         CoRE MOF  \cite{Chung2014}  &
         properties of metal-organic frameworks &
         5k &
         Open \\

         hMOF \cite{Wilmer2012}  &
         hypothetical metal-organic frameworks &
         140k &
         Open \\

         CSD \cite{groom_cambridge_2016}  &
         crystal structures &
         1 M &
         API \\

         ICSD \cite{Bergerhoff1983} & 
         crystal structure data for inorganic compounds & 
         180k & 
         Commercial  \\
         
         NOMAD \cite{noauthor_nomad_nodate}  &
         total energy calculations &
         50 M &
         Open \\
         
         AFLOW \cite{curtarolo_aflow:_2012,noauthor_aflow_nodate}  &
         material compounds; calculated properties &
         2.1M; 282M &
         Open \\
         
         OQMD \cite{kirklin_open_2015}  &
         total energy calculations &
         560 k &
         Open \\
         
         Materials Project \cite{jain_commentary:_2013} &
         inorganic compounds and computed properties  &
         87 k  &
         API \\

         Computational Materials Repository  \cite{landis_computational_2012,noauthor_projects_nodate} &
         inorganic compounds and computed properties  &
         varies &
         Open \\
         
         Pearson's \cite{noauthor_pearsons_nodate}  &
         crystal structures &
         319 k &
         Commercial \\

         HOPV \cite{lopez2016harvard} &
         experimental photovoltaic data from literature, QM calculations &
         350 k  & %
         Open \\ \hline 

         \multicolumn{4}{c}{ Journal articles  } \\ \hline
         Name &   Description &  Size & Availability \\  \hline 
         Crossref \cite{lammey_crossref_2015}   &
         journal article metadata  &
         107 M &
         Open \\
         
         PubMed \cite{pubmeddev_home_nodate} &
         biomedical citations &
         29 M &
         Open \\
         
         arXiv \cite{arxiv_download} & 
         arXiv articles (from many domains) & 
         1.6 M & 
         Open \\

         Wiley \cite{wiley_download}   &
         full articles &
         millions &
         API \\
         
         Elsevier \cite{elsevier_download}    &
         full articles &
         millions & 
         API \\

    \end{longtable}
    
\end{center}

Several factors have contributed to the greater wealth and accessibility of chemical databases that can be used to facilitate discovery. The first of these has to do with hardware: automated experimentation, especially that which is high-throughput in nature, has allowed us to generate data at a faster pace. Second, the successes of computational tools at leveraging these large quantities of data has created a self-catalyzing phenomenon: as the capabilities of tools are more frequently and widely demonstrated, the incentive to collect and curate large datasets that can be used by these tools has grown.

\begin{leftbar}
\challenge{Establish open access databases with standardized data representations} 
Time invested in the creation of open databases of molecules, materials, and processes can have an outsized  impact on discovery efforts that are able to make use of that data for supervised learning or screening. %

\end{leftbar}

Creating and maintaining these databases is not without its challenges. There's much to be done to capture and open-source the data generated by the scientific community. For cases where data must be protected by intellectual property agreements, we need software that can facilitate sharing between collaborators and guarantee privacy as needed \cite{Ekins2014, Hie2018}. Even standardizing representations can be challenging, particularly for polymeric materials with stochastic structures and process-dependent attributes \cite{audus_polymer_2017}.

Government funding agencies in the EU and US are starting to prioritize accessibility of research results to the broader community \cite{Tetko2016}. Further evolution of the open data policies will accelerate discovery through broader analysis of data (crowdsourcing discovery \cite{Ekins2010,Norman2011,Ekins2012}) and amalgamation of data for the purposes of machine learning.  Best practices among experimentalists must begin to include the storage of experimental details and results in searchable, electronic repositories.%

The data that exists in the literature that remains to be tapped by the curation efforts described above is vast. To access it, scientific researchers are gaining increasing interest in adapting information extraction techniques for use in chemistry \cite{krallinger2015chemdner,swain2016chemdataextractor, krallinger_information_2017, kim_materials_2017, zhai_improving_2019, zheng_text_2019}. Information extraction and natural language processing bring structure to unstructured data, e.g., published literature that presents information in text form. Methods have evolved from identifying co-occurrences of specifics words \cite{swanson_interactive_1997} to the formalization of domain-specific ontologies \cite{gomez-perez_ontologies_2013}, learned word embeddings \cite{tshitoyan_unsupervised_2019}, knowledge graphs and databases \cite{kim_materials_2017}, and causal models \cite{gyori_word_2017}. Learning from unstructured data presents additional challenges in terms of data set preparation and problem formulation, and is significantly less popular than working with pre-tabulated databases. Nevertheless, building knowledge graphs of chemical topics may eventually let us perform higher level reasoning \cite{battaglia_relational_2018} to identify and generalize from novel trends. 

\begin{leftbar}
\challenge{Address the inconsistent quality of existing data}
Existing datasets may not contain all of the information needed for a given prediction task (i.e., the input is underspecified or the schema is missing a metric of interest). Even when the right fields are present, there may be missing or misentered data from automated information extraction pipelines or manual entry.
\end{leftbar}

As \citeauthor{Williams2009} point out, data curation (which involves evaluating the accuracy of data stored in repositories) before the data is used to create a model or otherwise draw conclusions is very important: data submitted to the PDB is independently validated before it is added to the database, whereas data added to PubChem undergoes no prerequisite curation or validation \cite{Williams2009}. Lack of curation results in many problems associated with missing and/or misentered data. These issues plague databases including the PDB (misassigned electron density) and Reaxys (missing data). As analytical technology continues to improve, one can further ask how much we should bother relying on old data in lieu of generating new data that we trust more.

Existing database curation policies must account for the potential for error propagation and incorporate standardization procedures that can correct for errors when they arise \cite{Jaskolski2013,Berman2013}, for example by using ProsaII \cite{Sippl1993} to evaluate sequence-structure compatibility of PDB entries and identify errors \cite{Venclovas2004}. While the type of crowdsourcing error correction exemplified by \citeauthor{Venclovas2004} can be helpful, we argue that it shouldn't be relied upon \cite{Venclovas2004}; curators should preemptively establish policies to help identify, control, and prevent errors.

\subsubsection{Building empirical models}

Various statistical methods have been used for model-building for many years. Some of the most dramatic improvements to statistical learning have been in the area of machine learning. Machine learning is now the go-to for developing empirical models that describe nonlinear structure-function relationships that estimate the properties of new physical matter and serve as surrogate models for expensive calculations or experiments. These types of empirical models act as a roadmap for many discovery efforts, so improvements here significantly impact computer-aided discovery, even when the full workflow is not automated. Packages like scikit-learn, Tensorflow, and Pytorch have lowered the barrier for implementing empirical models, and chemistry-specific packages like DeepChem \cite{wu2018moleculenet} and ChemML \cite{haghighatlari_chemml:_2019} represent further attempts to streamline model training and deployment (with mixed success in adoption).

\begin{leftbar}
\challenge{Improve representations of molecules and materials} 
A wide array of strategies for representing molecules and materials for the sake of empirical modeling have been developed, but certain aspects have yet to be adequately addressed by existing representation methods. Further, it is difficult to know which representation method will perform best for a given modeling objective.
\end{leftbar}

In the wake of the 2012 ImageNet competition, in which a convolutional neural network dominated rule-based systems for image classification \cite{Krizhevsky2012}, there has been a shift in modeling philosophy to avoid human feature engineering, e.g., describing molecules by small numbers of expert descriptors, and instead \emph{learn} suitable representations \cite{hop_geometric_2018}. This is in part enabled by new network architectures, such as message passing networks particularly suited to embedding molecular structures \cite{duvenaud_convolutional_2015,kearnes_molecular_2016,gilmer_neural_2017}. There is no consensus as to when the aforementioned deep learning techniques should be applied over ``shallower'' learning techniques like RFs or SVMs with fixed representations \cite{korolev_graph_2019}; which method performs best is task-dependent and determined empirically \cite{wu2018moleculenet, yang_are_2019}, although some heuristics, e.g. regarding the fingerprint granularity needed for a particular materials modeling task, do exist \cite{Ramprasad2017}.  Use of molecular descriptors may make generalization to new inputs more predictable \cite{DerekTAhneman2018} but limits the space of relationships able to be described by presupposing that the descriptors contain all information needed for the prediction task. Further, selecting features for  low-dimensional descriptor-based representations requires expert-level domain knowledge.

In addition to a better understanding of why different techniques perform well in different settings, there is a need for the techniques themselves to better capture relevant information about input structures. Some common representations in empirical QSAR/QSPR modeling are listed in Table~\ref{tab:trends_representations}. However, there are several types of inputs that current representations are unable to describe adequately.  These include (a) polymers that are stochastically-generated ensembles of specific macromolecular structures, (b) heterogeneous materials with periodicity or order at multiple length scales, and (c) ``2.5D'' small molecules with defined stereochemistry but flexible 3D conformations. Descriptor-based representations serve as a catch-all, as they rely on experts to encode input molecules, materials, or structures as numerical objects.

\begin{table}[h]
    \centering
    \small
    \begin{tabular}{p{5cm}|p{10cm}}
        \hline 
        \bf Representation & \bf Description \\ \hline 
        Descriptors & Vector of calculated properties \\ 
        Fingerprints & Vector of presence/absence or count of structural features (many types) \\ 
        Coulomb matrices & Matrix of electrostatic interactions between nuclei \\ 
        Images & 2D line drawings of chemical structures \\
        SMILES & String defining small molecule connectivity (can be tokenized or adapted in various ways, e.g., SELFIES, DeepSMILES) \\
        FASTA & String for nucleotide or peptide sequences \\
        Graphs & 2D representation with connectivity information \\
        Voxels & Discretized 3D or 4D representation of molecules \\
        Spatial coordinates & 3D representation with explicit coordinates for every atom \\ 
        
        \hline 
    \end{tabular}
    \caption{Representations used in empirical QSAR/QSPR modeling. }
    \label{tab:trends_representations}
\end{table}

\begin{leftbar}
\challenge{Improve empirical modeling performance in low-data environments} 
Empirical modeling approaches must be validated on or extended to situations for which only tens of examples are available (e.g., a small number of hits from an experimental binding affinity assay).
\end{leftbar}

Defining a meaningful feature representation is especially important when data is limited \cite{ghiringhelli_big_2015, janet_resolving_2017}. Challenging discovery problems may be those for which little performance data is available and validation is expensive. For empirical models to be useful in these settings, they must be able to make reasonably accurate predictions with only tens of data points. QSAR/QSPR performance in low-data environments is understudied, with few papers explicitly examining low-data problems (e.g., fewer than 100 examples) \cite{altae-tran_low_2017,li_deep_2018,Zhang2018}. The amount of data ``required'' to train a model is dependent on the complexity of the task, the true (unknown) mathematical relationship between the input representation and output, the size of the domain over which predictions are made, and the coverage of the training set within that space.  %

\begin{leftbar}
\challenge{Incorporate physical invariance and equivariance properties}
By ensuring that models are only sensitive to meaningful differences in input representations, one can more effectively learn an input-output relationship without requiring data augmentation to also learn input-input invariance or equivariance. 
\end{leftbar}

One potential way to improve low-data performance and generalization ability is to embed physical invariance or equivariance properties into models. 
Consider a model built to predict a physical property from a molecular structure: molecular embeddings from message passing neural networks are inherently invariant to atom ordering. In contrast, embeddings calculated from tensor operations on Coulomb matrices are not invariant.  Sequence encoders using a SMILES string representation of a molecule have been shown to benefit from data augmentation strategies so the model can learn the chemical equivalence of multiple SMILES strings \cite{bjerrum_smiles_2017, bjerrum_improving_2018}. There are strong parallels to image recognition tasks, where one may want an object recognition model to be invariant to translation, rotation, and scale. When using 3D representations of molecules with explicit atomic coordinates, it is preferable to use embedding architectures that are inherently rotationally-invariant \cite{bartok_representing_2013, schutt_schnet:_2017} instead of relying on inefficient preprocessing steps of structure alignment \cite{zahrt_prediction_2019} and/or rotational enumeration \cite{ragoza_proteinligand_2017} for voxel representations, which still may lead to models that do not obey natural invariance or equivariance laws.

\begin{leftbar}
\challenge{Unify and utilize heterogeneous datasets} 
Vast quantities of unlabeled or labeled data can be used as baseline knowledge for pretraining empirical models or in a multitask setting when tasks are sufficiently related. 
\end{leftbar}

When human researchers approach a discovery task, they do so equipped with an intuition and knowledge base built by taking courses, reading papers, running experiments, and so on. In computational discovery workflows with machine learning-based QSAR modeling, algorithms tend to focus only on the exact property or task and make little use of prior knowledge; only via the input representation,  model architecture, and constraints on the search space is domain-specific knowledge embedded. Models are often trained from scratch on datasets that contain labeled (molecule, value) pairs. %

Such isolated applications of machine learning to QSAR/QSPR modeling can be  effective, but there is a potential benefit to multitask learning or transfer learning when predictions are sufficiently related \cite{dahl_multi-task_2014, ramsundar_massively_2015, fare_powerful_2018, Gupta2018, Merk2018}. \citeauthor{searls_data_2005} argues that drug discovery stands to benefit from integrating different datasets relating to various aspects of gene and protein functions \cite{searls_data_2005}. As a simple example, one can consider that the prediction of phenotypes from suppressing specific protein sequences might benefit from knowledge of protein structure, given the connection between protein sequence $\to$ structure $\to$ function. For some therapeutic targets, there are dozens of databases known to be relevant that have not been meaningfully integrated \cite{sundaramurthi_informatics_2012}. Large-scale pretraining is a more general technique that can be used to learn an application-agnostic atom- or molecule-level representation prior to refinement on the actual QSAR task \cite{goh_smiles2vec:_2017, goh_chemception:_2017, zhou_learning_2018, hu_pre-training_2019}. Performance on phenotypic assays has even been used directly as descriptors for molecules in other property prediction tasks \cite{sawada_target-based_2015}, as has heterogeneous data on drug, protein, and drug-protein interactions \cite{luo_network_2017}.

\begin{leftbar}
\challenge{Improve interpretability of machine learning models} 
Machine learning models are typically applied as black box predictors with some minimal degree of \emph{ex post facto} interpretation: analysis of descriptor importance, training example relevance, simplified decision trees, etc. Extracting explanations consistent with those used by human experts in the scientific literature requires the structure of the desired explanations to be considered and built into the modeling pipeline. 
\end{leftbar}

To the extent that existing autonomous discovery frameworks generate hypotheses that explain observations and interpret the results of experiments, they rarely do so in a way that is directly intelligible \emph{to humans}, limiting the expansion of scientific knowledge that is derived from a campaign. In connection with this, many of the case studies from Part 1 focus on discoveries that are readily physically observable--identifying a new molecule that is active against a protein target, or a new material that can be used to improve energy capture--rather than something more abstract, such as answering a particular scientific question. %
We can probe model understanding by enumerating predictions for different inputs, but these are human-defined experiments to answer human-defined hypotheses (e.g., querying a reaction prediction model with substrates across a homologous series). 
Standard approaches to evaluating descriptor importance still require careful control experiments to ensure that the explanations we extract are not spurious, even if they align with human intuition \cite{chuang_comment_2018}.  We again refer readers to ref.~\citenum{polishchuk_interpretation_2017} for a review of QSAR interpretability. 
Ref.~\citenum{roscher2019explainable} reviews additional aspects of explainable machine learning for scientific discovery.

\begin{figure}[h]
  \centering
  \includegraphics[width=16cm]{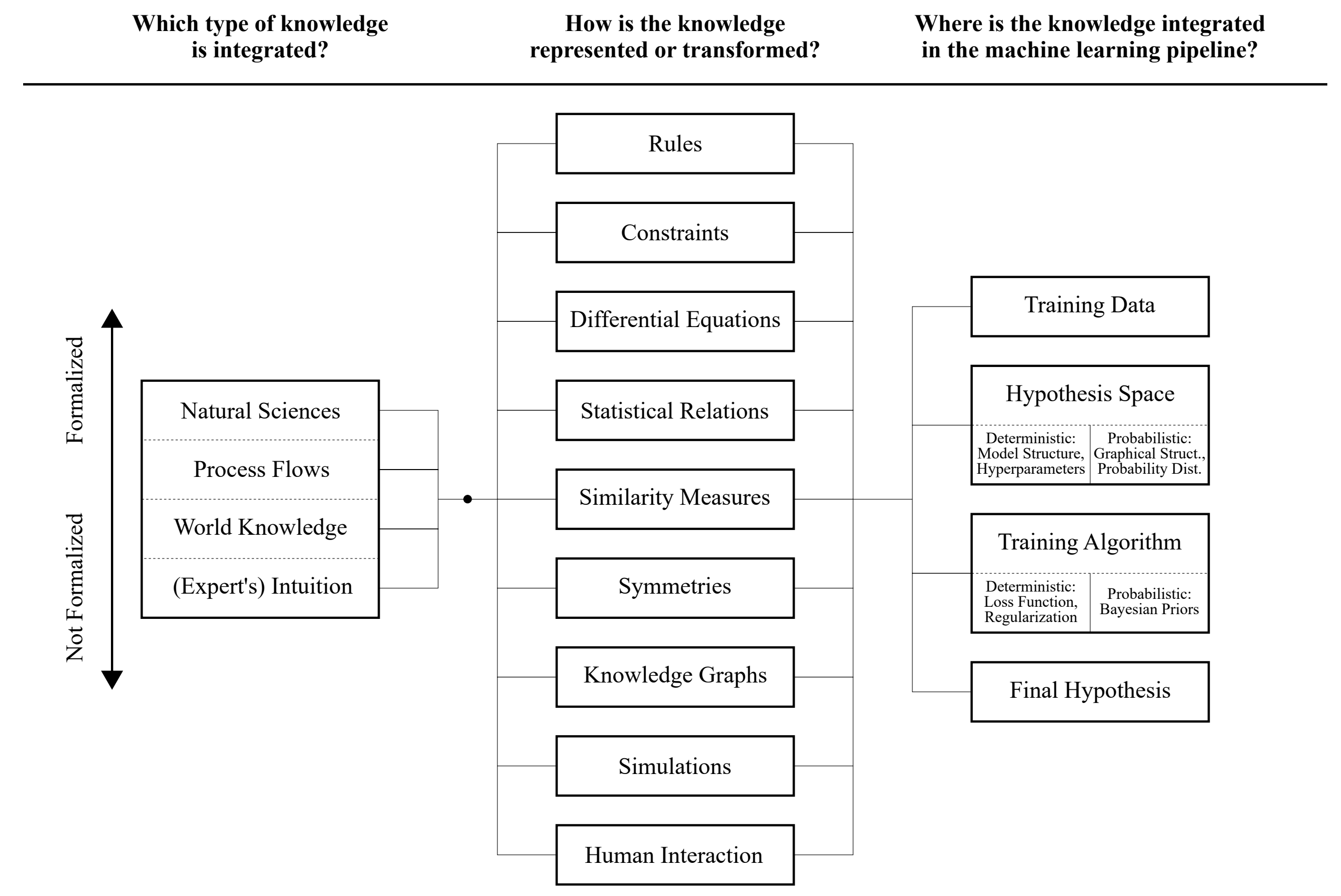}
  \caption{Taxonomy of \emph{informed machine learning} proposed by \citeauthor{von_rueden_informed_2019}. The incorporation of prior knowledge into machine learning modeling can take a number of forms. Figure reproduced from ref.~\citenum{von_rueden_informed_2019}.}
  \label{fig:von_rueden_taxonomy}
\end{figure}

Many challenges above can be approached by what \citeauthor{von_rueden_informed_2019} call \emph{informed machine learning}: ``the explicit incorporation of additional knowledge into machine learning models''. The taxonomy they propose is reproduced in Figure~\ref{fig:von_rueden_taxonomy}. In particular, several points relate to (a) the integration of natural sciences (laws) and intuition into representations and model architectures and (b) the integration of world knowledge through pretraining or multitask/transfer learning.

\subsection{Automated validation and feedback}

Iterating between hypothesis generation and validation can be fundamental to the discovery process. One often needs to collect new data to refine or prove/disprove hypotheses. Sufficiently advanced automation can compensate for bad predictions by quickly falsifying hypotheses and identifying false positives \cite{kitano_artificial_2016} (i.e., being ``fast to fail'' \cite{paul_how_2010}).  The last several decades have brought significant advances in automation of small-scale screening, synthesis, and characterization, which facilitates validation via physical experiments, as well as advances in software for faster and more robust computational validation. 

\subsubsection{Experimental validation}

Many of the case studies we present portray great strides in terms of the speed and scale of experimental validation. High-throughput and parallelized experimentation capabilities have been transformational in the biological space and increasingly are being imported into the chemistry space \cite{Shevlin2017}. The adoption of HTE has simplified screening broad design spaces for new information \cite{Santanilla2015, gesmundo_nanoscale_2018, Oliver2019}. Beyond brute-force experimentation, there are new \emph{types} of experiments to accelerate the rate of data generation and hypothesis validation. These include split-and-pool techniques and other combinatorial methods to study multiple candidates simultaneously \cite{Green2013, oreilly_evolution_2017,troshin_snap_2017}.

\begin{leftbar}
\challenge{Expand the scope of automatable experiments}
Whether an iterative discovery problem’s hypotheses can be autonomously validated depends on whether the requisite experiments are amenable to automation.
\end{leftbar}

If we are optimizing a complex design objective, such as in small molecule drug discovery, we benefit from having access to a large search space.
Many syntheses and assays are compatible with a well-plate format and are routinely automated (e.g., Adam \cite{king_functional_2004} and Eve \cite{williams_cheaper_2015}). Moving plates, aspirating/dispensing liquid, and heating/stirring are all routine tasks for automated platforms. Experiments requiring more complex operations may still be automatable, but require custom platforms, e.g., for the growth and characterization of nanotubes by ARES \cite{nikolaev_discovery_2014} or deposition and characterization of thin films by Ada \cite{macleod_self-driving_2019}.  Dispensing and metering of solids is important for many applications but is challenging at milligram scales, though new strategies are emerging that may decrease the precision required for dosing solid reagents \cite{wang_high_nodate}. Indeed, the set of automatable experiments is ever-increasing, but a universal chemical synthesizer \cite{lowe_derek_automated_2018} remains elusive. The result of this gap is that design spaces may be constrained not only through prior knowledge (an intentional and useful narrowing of the space), but also \emph{limited} by the capabilities of the automated hardware available. Characterizing the structure of physical matter is increasingly routine, but our ability to measure complex functions and connecting those back to structure remains limited.  \citeauthor{Oliver2019} list several useful polymer characterization methods that have eluded full automation, such as differential scanning calorimetry and thermogravimetric analysis \cite{Oliver2019}.

\begin{leftbar}
\challenge{Facilitate integration through systems engineering} %
Scheduling, performing, and analyzing experiments can involve coordinating tasks between several independent pieces of hardware/software that must be physically and programmatically linked.
\end{leftbar}

\begin{figure}[h]
  \centering
  \includegraphics[width=7cm]{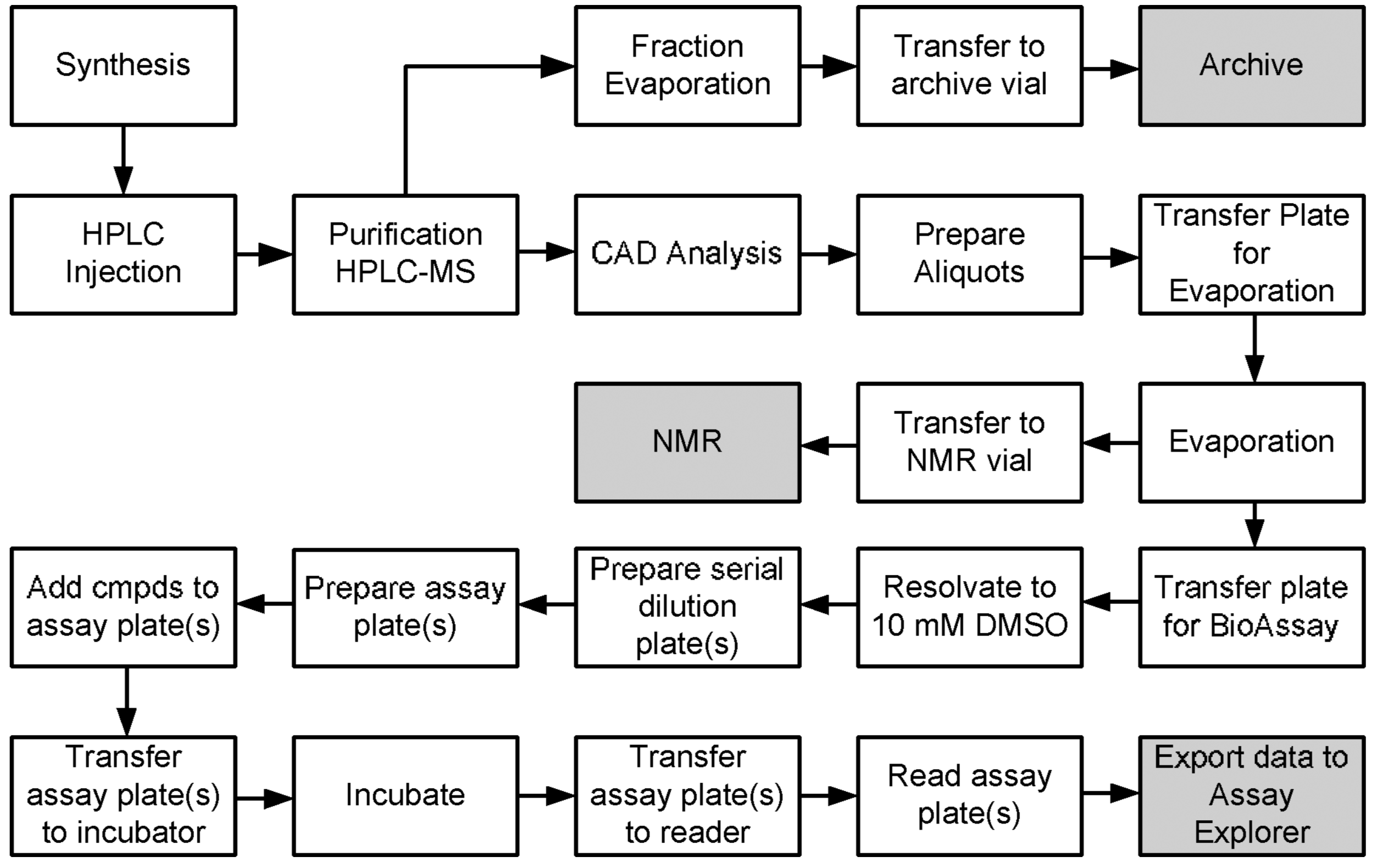}
  \caption{The workflow of automated synthesis, purification, and testing requires the scheduling of many independent operations handled by different pieces of hardware and software. Figure reproduced from \citeauthor{baranczak_integrated_2017} \cite{baranczak_integrated_2017}.}
  \label{fig:baranczak_integrated_2017_workflow}
\end{figure}

Expanding the scope of experimental platforms may require the integration of  independent pieces of equipment at both the hardware and software level. The wide variety of necessary tasks (scheduling, error-handling, etc.) means that designing control systems for such highly-integrated platforms is an enormously complex task \cite{Pan2019}. As a result, developing software for integration of an experimental platform \cite{baranczak_integrated_2017} (Figure~\ref{fig:baranczak_integrated_2017_workflow}) can be a large contributor to the cost. The lack of a standard API and command set between different hardware providers means that each requires its own driver and software wrapper; this is particularly true for analytical equipment, which even lacks standardization in file formats for measured data. Programs like OVERLORD and \citeauthor{roch_chemos:_2018}'s ChemOS \cite{roch_chemos:_2018} are attempts to create higher-level controllers. Throughput-matching in sequential workflows is a challenge in general, requiring a plan for ``parking'' (and perhaps stabilizing) samples in the event of a bottleneck downstream. These practical issues must be resolved to benefit from increased integration and the ability to generate data.

\begin{leftbar}
\challenge{Automate the planning of multistep chemical syntheses}
Many discovery tasks involve proposing new chemical matter; approaching these tasks with autonomous systems requires the ability to synthesize novel compounds on-demand. 
\end{leftbar}

\begin{figure}[h]
  \centering
  \includegraphics[width=14cm]{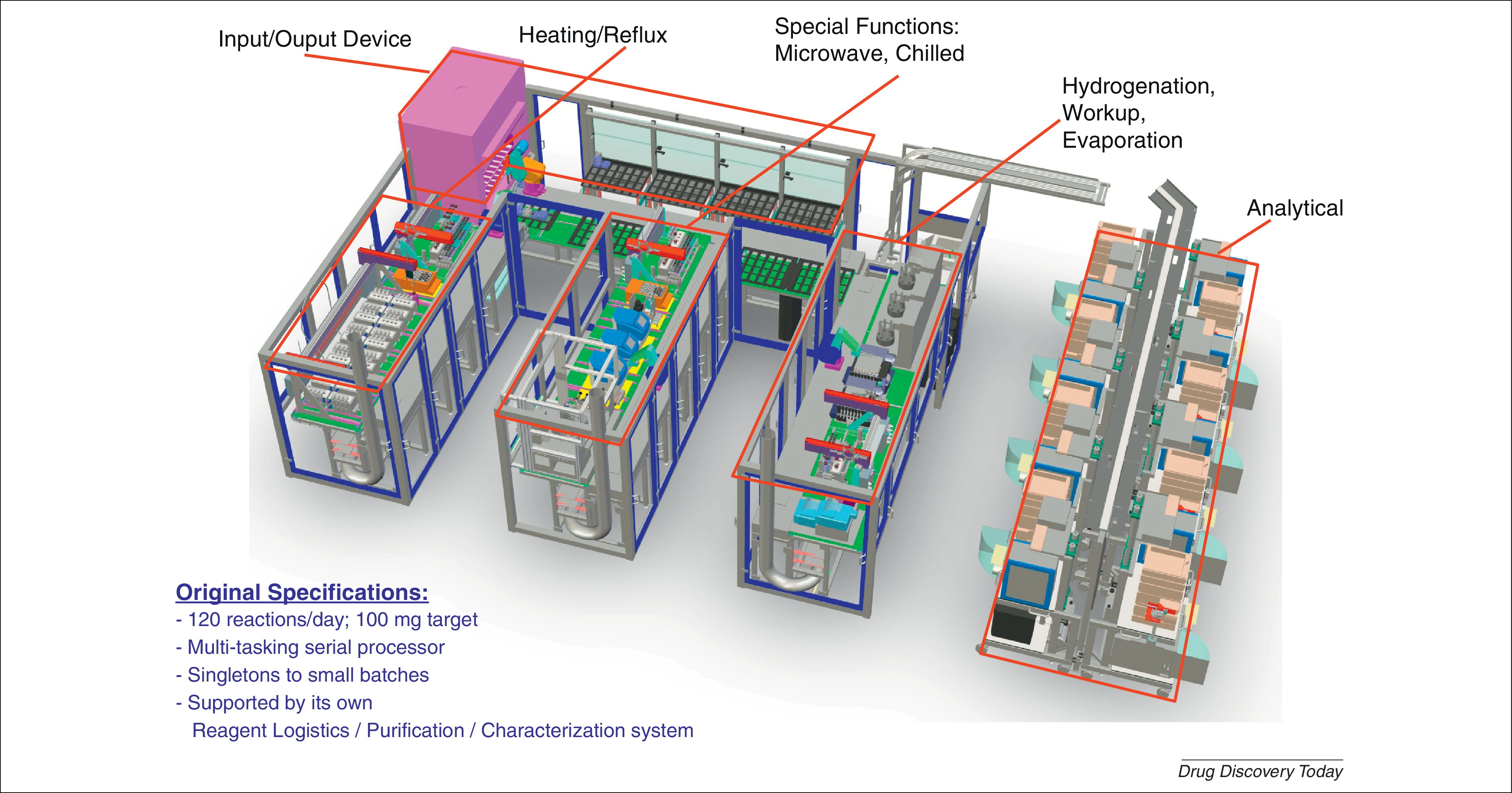}
  \caption{Rendering of Eli Lilly's first generation Automated Synthesis Laboratory (ASL) for automated synthesis and purification. Reproduced from ref.~\citenum{godfrey_remote-controlled_2013}. }
  \label{fig:godfrey_asl}
\end{figure}

A particularly challenging class of experiments is on-demand synthesis. 
The primary methodological/intellectual challenge for general purpose automated synthesis is the \emph{planning} of processes--designing multistep synthetic routes using available building blocks; selecting conditions for each reaction step including quantities, temperature, and time; and automating  intermediate and final purifications.
If reduced to stirring, heating, and fluid transfer operations, chemical syntheses are straightforward to automate \cite{machida_development_2010, steiner_organic_2018, jiang_integrated_2019}, and robotic platforms (Figure~\ref{fig:godfrey_asl}) are capable of executing a series of process steps if those steps are precisely planned \cite{godfrey_remote-controlled_2013, baranczak_integrated_2017}. However, current CASP tools are unable to make directly implementable recommendations with this level of precision. 

There are two diverging philosophies of how to approach automated synthesis: (a) the development of general-purpose machines able to carry out most chemical reactions, or (b) the development of specialized machines to perform a few general-purpose reactions that are still able to produce most molecules. The references in the preceding paragraph follow the former approach.
Burke and co-workers have advocated for the latter and propose using advanced MIDA boronate building blocks and a single reaction/purification strategy to simplify process design \cite{li_synthesis_2015}. Peptide and nucleic acid synthesizers exemplify this notion of automating a small number of chemical transformations to produce candidates within a vast design space. 

\subsubsection{Computational validation}

Many discoveries can be validated with high confidence through computational techniques alone. Where applicable, this can be extremely advantageous. This is because the logistics of the alternative (physical experiments) may be much more complex, e.g., relying on  access to a large chemical inventory (of candidates or as precursors) to perform validation experiments within a large design space. An emblematic example of discoveries that can be validated through computation alone is that of physical matter whose desired function can be reliably estimated with first principles calculations.  

\begin{leftbar}
\challenge{Accelerate code/software used for computational validation}
Just as in physical experiments, there are practical challenges in computational experiments related to the throughput of high fidelity calculations.
\end{leftbar}

A unique feature of computational validation is the well-recognized tradeoff between speed and accuracy. Consider \citeauthor{lyu_ultra-large_2019}'s virtual screen of 170 million compounds to identify binders to two protein targets through rigid-body docking \cite{lyu_ultra-large_2019}. The computational pipeline was fast--requiring about one second per compound--only requiring $\approx$ 50,000 core-hours in total, and successfully yielded dozens of novel chemotypes and a few sub-nanomolar binders. While rigid-body docking is not as robust as, say, explicit molecular dynamics, its predictions still correlated with experimental binding and generated useful candidate molecules.  An earlier study by \citeauthor{gomez-bombarelli_design_2016} used time-dependent DFT to evaluate a total of 400,000 candidate OLED molecules selected from a pool of 1.6 million enumerated candidates; the computational cost for this study was roughly 13 million core-hours \cite{gomez-bombarelli_design_2016}. There are billions upon billions of molecules in enumerated synthetically-accessible chemical space. Given our desire to work with broad design spaces, there is a need for faster workflows that can conduct large-scale computational validation. 

One strategy to accelerate computational validation is to create surrogate models of first-principles calculations \cite{gomez-bombarelli_design_2016, janet_accelerating_2018, smith_ani-1:_2017}. Predictions made using surrogate machine learning models regressed to measured or simulated data almost always carry greater uncertainty than the original experiment, and therefore confirming the final discovery is especially reliant on higher-fidelity, often physical validation. An orthogonal strategy is to accelerate the calculations themselves without sacrificing accuracy. Many computational chemistry and molecular modeling packages have been developed to take advantage of hardware acceleration on GPUs \cite{stone_accelerating_2007, ufimtsev_graphical_2008, stone_gpu-accelerated_2010}, FPGAs \cite{yang_fully_2019}, and even ASICs \cite{shaw_anton_2008}.

\begin{leftbar}
\challenge{Broaden capabilites / applicability of first-principles calculations}
Many properties of interest cannot be simulated accurately, forcing us to rely on experimental validation.
\end{leftbar}

Expanding the scope of what can be accurately modeled would open up additional applications for purely computational autonomous workflows. There are some tasks for which computational solutions exist but could be improved, including binding prediction through docking, reactive force field modeling, transition state searching, conformer generation, solvation energy prediction, and crystal structure prediction. There are other tasks with even fewer satisfactory approaches, including long timescale molecular dynamics and multiscale modeling in materials. Some grand challenges in computational chemistry are discussed in refs.~\citenum{hofer_macromolecules_2013} and \citenum{grimme_computational_2018}. %

\subsubsection{Shared challenges}

\begin{leftbar}
\challenge{Ensure that validation reflects the real application}
Computational or experimental validation that lends itself to automation is often a proxy for a more expensive evaluation. If the proxy and the true metric are misaligned, an autonomous platform will not be able to generate any useful results.
\end{leftbar}

Ideally, there would be perfect alignment between the approaches to validation compatible with an autonomous system and the real task at hand. This is impossible for tasks like drug discovery, where imperfect \emph{in vitro} assays are virtually required before evaluating \emph{in vivo} performance during preclinical development. For other tasks, assays are simplified for the sake of automation or cost, e.g., Ada's measurement of optoelectronic properties of a thin film as a proxy for hole mobility as a proxy of the efficiency of a multicomponent solar cell \cite{macleod_self-driving_2019}. Assays used for validation in autonomous systems do not necessarily need to be high throughput, just high fidelity and automatable. Avoiding false results, especially false negatives in design spaces where positive hits are sparse such as in the discovery of ligands that bind strongly to a specific protein, is critical \cite{Malo2006}.%

\begin{leftbar}
\challenge{Lower the cost of automated validation}
Relatively few things can be automated cheaply. This is especially true for problems requiring complex experimental procedures, e.g., multi-step chemical synthesis. 
\end{leftbar}
While the equipment needed for a basic HTE setup is becoming increasingly accessible and compatible with the budget of many academic research groups \cite{baker_academic_2010, allen_power_2019},  we must increase the complexity of the automated platforms that are used for validation in order to increase the complexity of problems that can be addressed by autonomous workflows. Autonomous systems need not be high-throughput in nature, but, as we have mentioned several times throughout this review, accelerating search to facilitate exploration of ever-broader design spaces that we cannot explore manually should be one of the key goals/outcomes of development of these types of platforms. 
As the community begins to undertake this challenge, it's imperative that we pay attention to affordability, lest we discourage/inhibit adoption. Homegrown systems can be made inexpensively through integration of common hardware components and open-source languages for control \cite{OBrien2017, steiner_organic_2018}.  Miniaturization reduces material consumption costs, but can complicate system fabrication and maintenance. The decision to automate a workflow should be the result of a holistic evaluation of return on investment (ROI)  \cite{Pan2019}. 

The costs of computational assays are less of an impediment to autonomous discovery than experimental assays, given the accessibility of large-scale compute. Improving their accuracy is more of a priority. For example, the docking method used by \citeauthor{lyu_ultra-large_2019} was sufficiently inexpensive to screen millions of compounds and obtain results that correlate with experimental binding affinity, but the majority of high scoring compounds are false positives and the differentiation of top candidates is poor \cite{lyu_ultra-large_2019,su_comparative_2019}.

\begin{leftbar}
\challenge{Combine newly acquired data with prior literature data}
Predictive models trained on existing data reflect beliefs about structure-property landscapes; when new data is acquired, that belief must be updated, preferably in a manner that reflects the relative confidence of the data sources.
\end{leftbar}

A fundamental question yet to be addressed in studies combining data mining with automated validation is the following: how should new data acquired through experimental/computational validation be used to update models pretrained on literature data? The quintessential workflow for non-iterative data-driven discovery of physical matter includes (a) regressing a structure-property dataset, (b) proposing a new molecule, material, or device, and (c) validating the prediction for a small number of those predictions. Incorporating this new data into the model should account for the fact that the new data may be generated under more controlled conditions or may be higher fidelity than the literature data.

The nature of existing data can be different from what is newly acquired. For example, tabulated reaction data is available at the level of chemical species, temperature, time, intended major product, and yield. In the lab, we will know the conditions quantitatively (e.g., concentrations, order of addition), will have the opportunity to record additional factors (e.g., ambient temperature, humidity), and may be able to measure additional endpoints (e.g., identify side products). However, while we can more thoroughly evaluate different reaction conditions than what has been previously reported, the diversity of substrates reported in the literature exceeds what is practical to have in-stock in any one laboratory; we must figure out how to meaningfully integrate the two. For discovery tasks that aim to optimize physical matter with standardized assays, where databases contain exactly what we would calculate or measure, this notion of complementarity is less applicable.

\subsection{Selection of experiments for validation and feedback}

Excellent foundational work in statistics on (iterative) optimal experimental design strategies has been adapted to the domain of chemistry. Although iterative strategies often depend on manually-designed initializations and constrained search spaces, algorithms can be given the freedom to make decisions about which hypotheses to test. This flexibility makes iterative strategies inherently more relevant to autonomous discovery than noniterative ones.

A variety of algorithms exist for efficiently navigating design spaces and/or compound libraries (virtual or otherwise). Broadly speaking, these can be categorized as model-free--black box optimizations, including evolutionary algorithms (EAs)--or model-based--using surrogate models for predicting performance and/or model uncertainty. The latter category includes uncertainty-guided experimental selection where an acquisition function quantifies how useful a new experiment would be \cite{settles2012active}; ref.~\citenum{frazier2018tutorial} provides a tutorial on Bayesian optimization.

\begin{leftbar}
\challenge{Quantify model uncertainty and domain of applicability} 
Active learning strategies are crucially dependent on quantifying uncertainty; doing so reliably in QSAR/QSPR modeling remains elusive, and current strategies cannot anticipate structure-activity cliffs or other rough features. 
\end{leftbar}

Accurate uncertainty quantification drives discovery by drawing attention to underexplored areas of a design space and helping to triage experiments, e.g., in combination with Bayesian optimization \cite{janet_designing_2019}. Statistical and probabilistic frameworks can account for uncertainty when analyzing data and selecting new experiments \cite{ghahramani_probabilistic_2015, Ueno2016, Peterson2017, hase_phoenics:_2018, janet_designing_2019}, but we must be able to meaningfully estimate our uncertainty to use them. Common frequentist methods for estimating uncertainty include model ensembling \cite{cortes-ciriano_deep_2019} and Monte Carlo (MC) dropout \cite{gal_dropout_nodate}; various Bayesian approaches like the use of Gaussian process models have been used as well \cite{hase_phoenics:_2018, simm_error-controlled_2018}. Not only is it difficult to generate meaningful outcomes with these methods, but also they tend to be computationally expensive (although MC dropout is generally less so than the others). In QSAR/QSPR, one often tries to define a domain of applicability (DOA) as a coarser version of uncertainty, where the DOA can be thought of as the input space for which the prediction and uncertainty estimation is meaningful \cite{tropsha_predictive_2007,Tropsha2010,toplak_assessment_2014}. 

There is little to no agreement on the correct way to estimate epistemic uncertainty (as opposed to aleatoric uncertainty, which is that which arises from measurement noise). In drug discovery, activity cliffs \cite{stumpfe_exploring_2012}--sharp changes in binding affinity resulting from minor structural changes--are especially troublesome and call into question any attempt to directly connect structural similarity to functional similarity \cite{bajorath_representation_2017, liu_molecular_2019}. Even functional descriptor-based representations are unlikely to capture all salient features. Implicit or explicit assumptions must be made  when choosing a representational and modeling technique, for example choosing an appropriate kernel and a prior on (or a hyperparameter controlling) the smoothness of the landscape in a Gaussian processes model \cite{obrezanova_gaussian_2007}.

\begin{leftbar}
\challenge{Quantify the tradeoff between experimental difficulty and information gain}
Experiment selection criteria should be able to account for the difficulty of an experiment, i.e., employ cost-sensitive active learning.
\end{leftbar}

Experiment selection methods rarely account for the \emph{cost} of an experiment in any quantitative way. Separately, experiment selection is occasionally biased based on factors that are irrelevant to the hypothesis. 
If proposed experiments require the synthesis of several molecules (e.g., a compound library designed during lead optimization), an expert chemist will generally select those they determine to be easily synthesized, rather than those that are most informative. One must ask if it is worth spending weeks making a single compound that maximizes the expected improvement or if there is a small analogue library that is easier to synthesize that, collectively, offers a similar probability of improvement. In this setting, there will almost always be a tradeoff between data that is fast and inexpensive to acquire and data that is most useful for the discovery. 
Understanding that tradeoff is essential for autonomous systems where experiments can have very different costs (e.g., selecting molecules to be synthesized) or likelihoods of success (e.g., electronic structure simulations prone to failure) in contrast to where experiments have similar costs (e.g., selecting virtual molecules for rigid-body docking). The situation becomes more complex for batched optimizations where, e.g., the cost of synthesizing 96 molecules in a parallel well-plate format is not merely the sum of their individual costs, but depends on overlap in the precursors and reaction conditions they employ.

\citeauthor{williams_cheaper_2015} provide one example of how to roughly quantify the value of active learning-based screening for Eve \cite{williams_cheaper_2015}. It is easy to imagine how one might augment this framework to account for cost as part of the experiment selection process. However, the utility calculation heuristics used by \citeauthor{williams_cheaper_2015} would need to be substantially improved in order to be usefully applied to cases where the cost of experiments vary, which is the interesting setting here. To date, the experiments able to be conducted by a given automated or autonomous workflow are of comparable cost; the decision about whether that cost is reasonable is made by the human designer of the platform. 
The term in experimental design for this is \emph{cost-sensitive} active learning \cite{donmez_proactive_2008}.

\begin{leftbar}
\challenge{Define discovery goals at a higher level} %
The ability of an automated system to make surprising or significant discoveries relies upon its ability to extrapolate and explore beyond what is known. This could be encouraged by defining broader objectives than what is currently done.
\end{leftbar}

\begin{figure}[h]
  \centering
  \includegraphics[width=12cm]{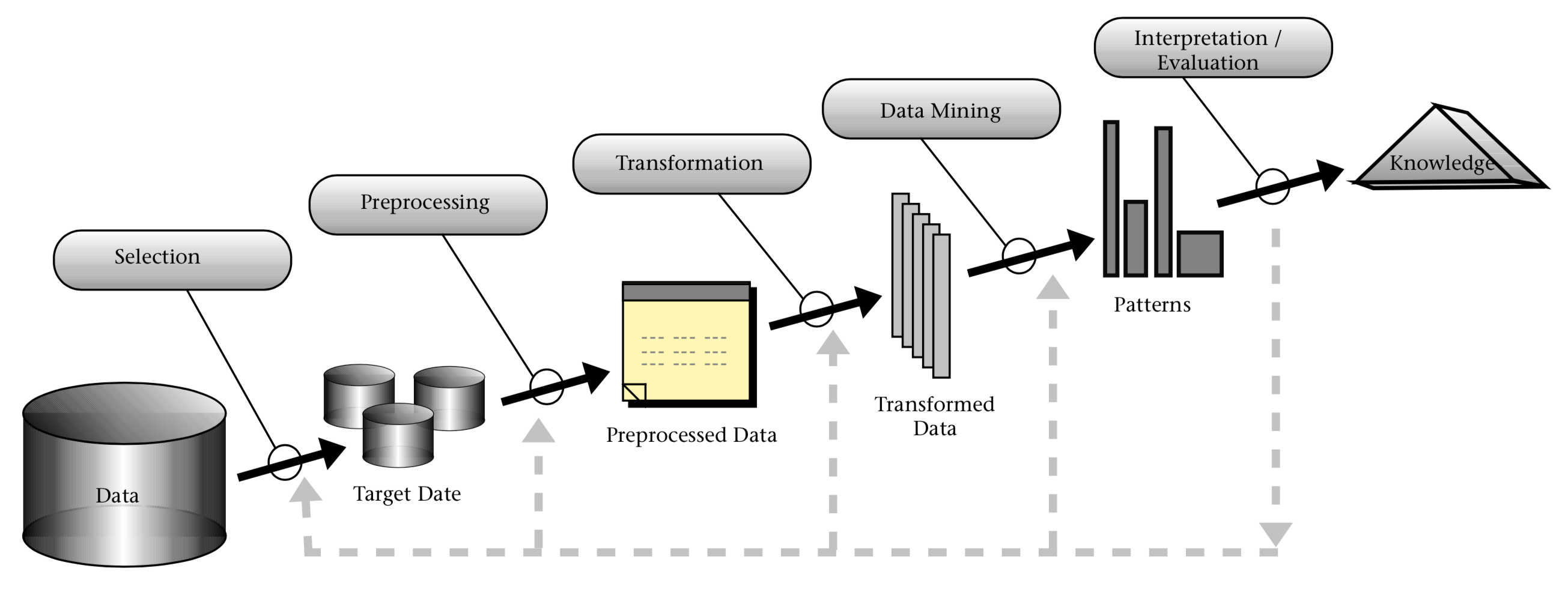}
  \caption{Overview of the process of knowledge discovery from databases. Figure reproduced from \citeauthor{fayyad_data_1996} \cite{fayyad_data_1996}.}
  \label{fig:fayyad_data_1996}
\end{figure}

In current data analyses, the structure of hypotheses tend to be prescribed: a mathematical function relating an expert-selected input to an expert-selected output, a correlative measure between two chemical terms, a causal model that describes a sequence of events. Ideally, we would be able to generate hypotheses from complex datasets in a more open-ended fashion where do not have to know exactly what we are looking for. Techniques in knowledge discovery \cite{fayyad_data_1996}, unsupervised learning \cite{bengio_unsupervised_2012}, and novelty detection \cite{pimentel_review_2014} are intended for just that purpose and may present a path toward more open-ended generation of scientific hypotheses (Figure~\ref{fig:fayyad_data_1996}). 

Experimental design can also be given greater flexibility by defining broad goals for discovery (performance, novelty, etc.) and using computational frameworks to learn tradeoffs in reaching those goals, e.g., through reinforcement learning. 
Consider the goal of compound selection during a drug discovery campaign: to identify a molecule that ends up being suitable for clinical trials. In the earlier information-gathering stages, we don't necessarily need to select the highest performing compounds, just the ones that provide information that lets us eventually identify them (i.e., a future reward). More generally, the experiments proposed for validation and feedback in a discovery campaign should be selected to achieve a higher-order goal (\emph{eventually}, finding the best candidate) rather than a narrow objective (maximizing performance within a designed compound library).

Open-ended inference is a general challenge in deep learning \cite{marcus_deep_2018}, as is achieving what we would call creativity in hypothesis generation \cite{boden_creativity_1998}. 
At some level, in order to apply optimization strategies for experimental design or analysis, the goal of a discovery search must be reducible to a scalar objective function. 
We should strive to develop techniques for guided extrapolation toward the challenging-to-quantify goals that the field  has used when defining discovery: novelty, interestingness, intelligibility, and utility.

\subsubsection{Proposing molecules and materials}

\begin{figure}[h]
  \centering
  \includegraphics[width=\linewidth]{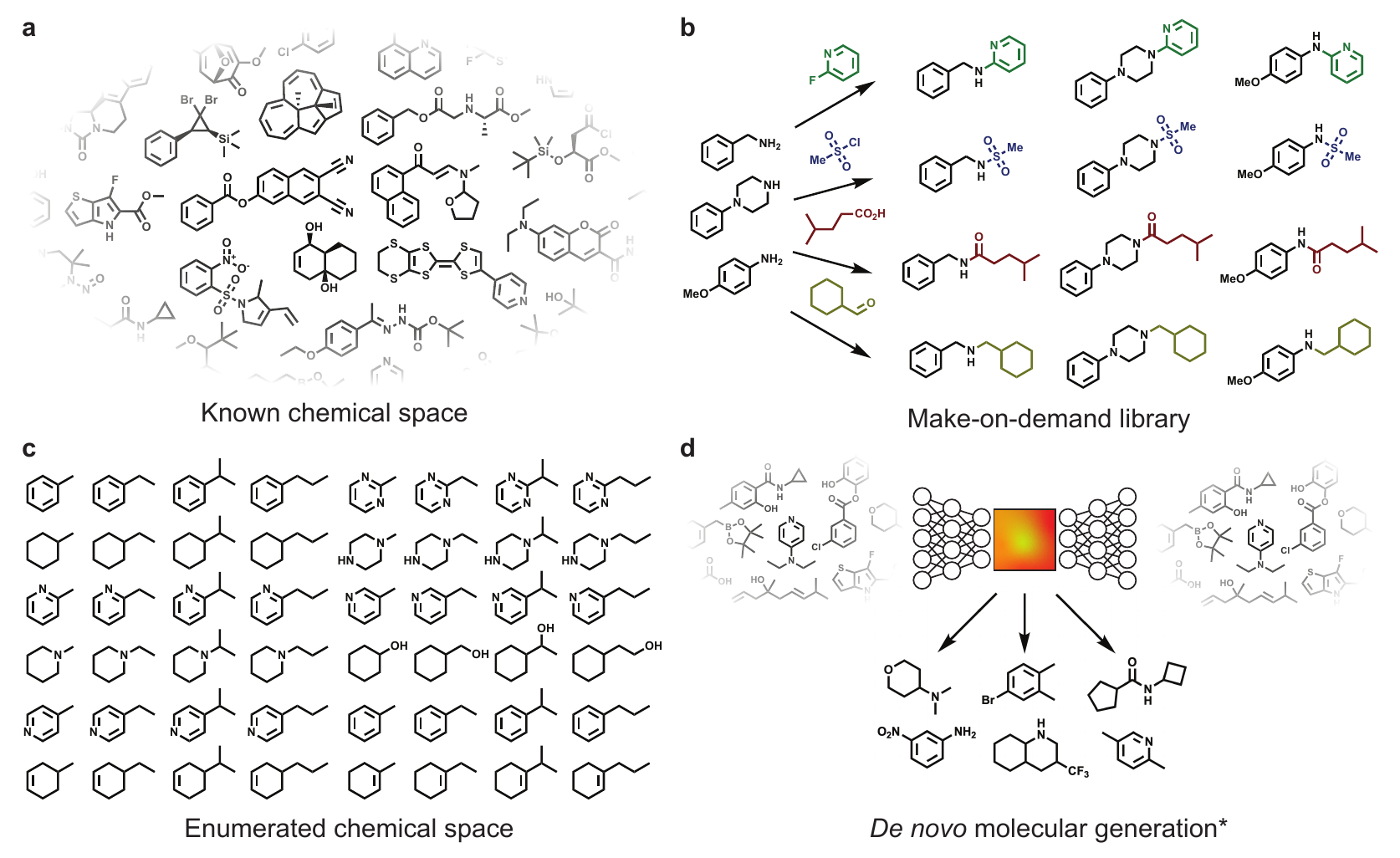}
  \caption{Common sources of molecules from which to select those that fulfill some design objective. Molecules can be selected from (a) a fixed, known chemical space, (b) a make-on-demand library of synthesizable compounds, (c) an enumerated library (via systematic enumeration or evolutionary methods), and (d) molecules proposed \emph{de novo} from a generative model. *An autoencoder architecture is shown as a representative type of generative model.} %
  \label{fig:proposing_molecules}
  \end{figure}

\begin{figure}[h]
  \centering
  \includegraphics[width=12cm]{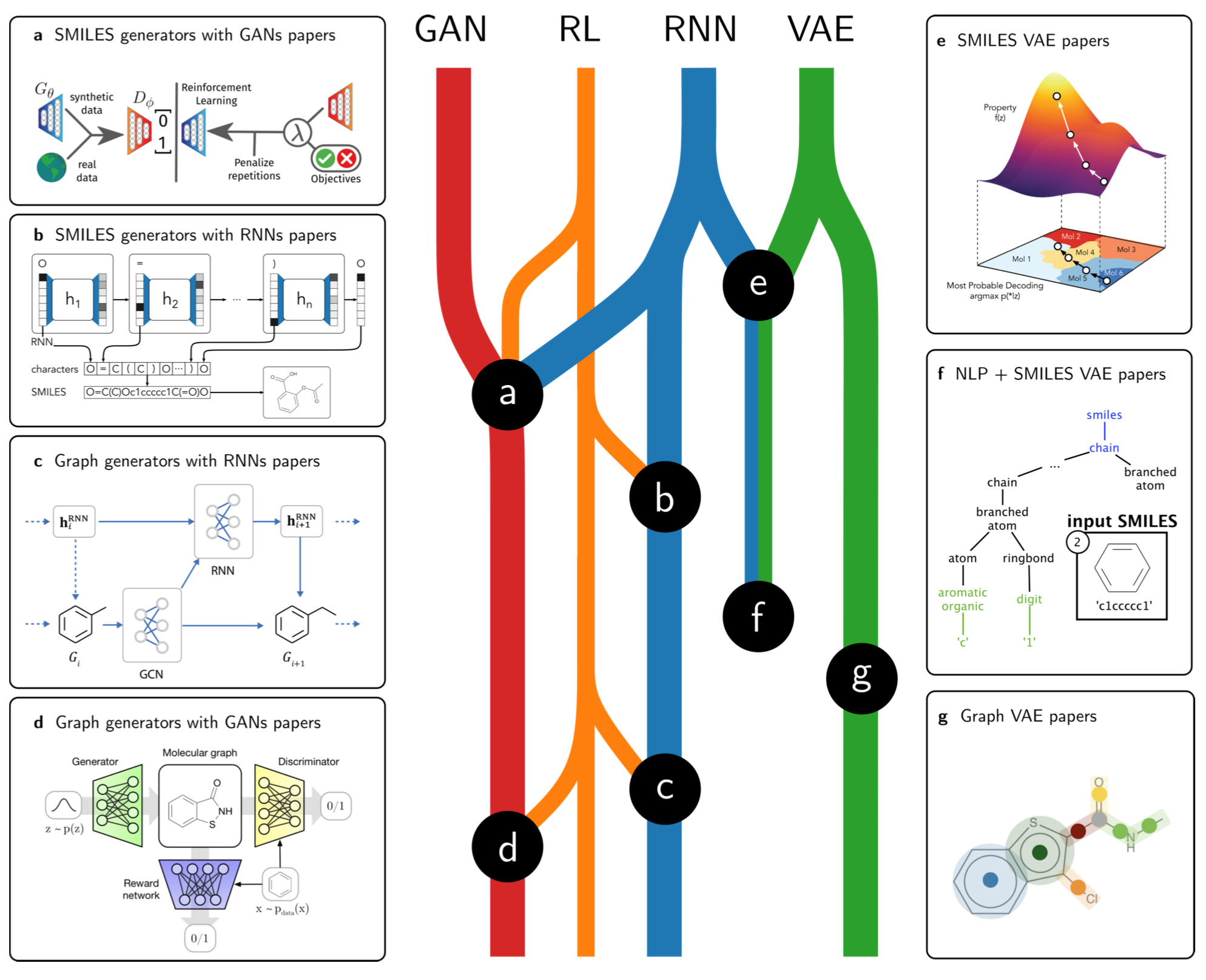}
  \caption{\citeauthor{schwalbe-koda_generative_2019}'s timeline of generative model development for molecules (top to bottom). Figure reproduced from ref.~\citenum{schwalbe-koda_generative_2019}. Figure subparts reproduced from refs. \citenum{Gomez-Bombarelli2018}, \citenum{guimaraes_objective-reinforced_2017}, \citenum{kusner_grammar_2017}, \citenum{jin_junction_2018}, \citenum{de_cao_molgan:_2018}, and \citenum{li_learning_2018}. } 
  \label{fig:schwalbe_2019_generative}
\end{figure}

Strategies for selecting molecules and materials for validation in discovery workflows are worth additional discussion (Figure~\ref{fig:proposing_molecules}). Iterative strategies of the sort described above apply here, with active learning being useful for selecting compounds from a fixed virtual library and evolutionary/generative algorithms being useful for designing molecules on-the-fly. 
Generative models are a particularly attractive way to design molecules and materials with minimal human input, biased only by knowledge of the chemical space on which they are trained  (Figure~\ref{fig:schwalbe_2019_generative}) \cite{sanchez-lengeling_inverse_2018, schwalbe-koda_generative_2019, elton_deep_2019}. 

\begin{leftbar}
\challenge{Bias generative models towards synthetic accessibility}
Compared to fixed virtual libraries, a shortcoming of generative models is that the molecules or materials they propose may not be easily realizable. 
\end{leftbar}

Algorithms that can leverage existing data to suggest promising, as-yet-untested possibilities exist, but these do not yet function on the level of a human scientist in part because they do not understand what experiments are possible.  Generative models can concoct new molecules in some abstract latent space, but simplistic measures of synthesizability \cite{ertl_estimation_2009, coley_scscore:_2018} are not enough to steer the models toward accessible chemical space. Make-on-demand virtual libraries provide a distinct advantage in that one is more confident proposed molecule can be made in short timeframe. Achieving that same confidence will be essential for the adoption of \emph{de novo} methods, some of which are beginning to combine molecular generation and virtual enumeration \cite{bradshaw_model_2019}. Some applications of generative models, like to peptide design, do not suffer from this limitation as, to a first approximation, most peptides are equally synthesizable.

\begin{leftbar}
\challenge{Benchmark problems for molecular generative models}
The current evaluations for generative models do not reflect the complexity of real discovery problems. 
\end{leftbar}

The explosion of techniques for molecular generation has outpaced our ability to meaningfully assess their performance. A metric introduced early on as a proxy objective is the ``penalized logP'' metric for molecular optimization. While not used for any actual discovery efforts, a heuristic function of estimated logP, synthetic accessibility, and a penalty for rings larger than 6 atoms was introduced for (and continues to be used for) benchmarking. This metric bears little resemblance to any multiobjective function one would use in practice. Only recently have more systematic benchmarks been introduced to cover a wider range of learning objectives: either maximizing a scalar objective or learning to mimic a distribution of training molecules. Two  frameworks for such model comparisons include GuacaMol \cite{brown_guacamol:_2018} and MOSES \cite{Polykovskiy2018moses}. However, these do not consider the number of function evaluations required by each method and still represent simplistic goals. Optimization goals that better reflect the complexity of real discovery tasks might include binding or selectivity as predicted by docking scores \cite{aumentado-armstrong_latent_2018}

\subsection{Evaluation}

\begin{leftbar}
\challenge{Demonstrate extrapolative power of predictive models} 
If the ultimate goal of computer-aided discovery is to generate new scientific knowledge, extrapolation beyond what is known is a necessity.
\end{leftbar}

The majority of approaches to automated discovery of physical matter rely on predictive models to guide the selection of experiments. The most effective models to facilitate this process will be able to at least partially extrapolate from our current knowledge to new chemical matter, eliminating the need for brute force experimentation. This extrapolative power--the ability of QSAR/QSPR models to generalize to design spaces they have not been trained on--should be prioritized as an  evaluation metrics during model development.  The potential for algorithms to guide us toward novel areas of chemical or reactivity space was emphasized in a recent review by \citeauthor{Gromski2019} \cite{Gromski2019}.

\begin{leftbar}
\challenge{Demonstrate design-make-test beyond proof-of-concept} %
All studies to-date that demonstrate closed-loop design, make, test cycles have been proof-of-concepts limited to narrow search spaces, severely limiting their practical utility. 
\end{leftbar}

A compelling demonstration of autonomous discovery in chemistry would be the closed-loop design, synthesis, and testing of new molecules to optimize a certain property of interest or build a structure-property model. There has not been much progress since early proof-of-concept studies that could access only a limited chemical space \cite{desai_rapid_2013, czechtizky_integrated_2013} despite significant advances in the requisite areas of molecular design, CASP, and automated synthesis. These constraints on the design space to ensure compatibility with automated validation prevent us from addressing many interesting questions and optimization objectives. \citeauthor{chow_streamlining_2018} describe several case studies where certain steps in the drug discovery have been integrated with each other for increased efficiency, but acknowledge--as others have--that \emph{all} stages must be automated and integrated for maximal efficiency \cite{chow_streamlining_2018, nicolaou_idea2data:_2019, saikin_closed-loop_2019}.

\begin{leftbar}
\challenge{Develop benchmark problems for discovery}
Developing methods for autonomous discovery would benefit from a ``sandbox'' that doesn't eliminate all of the complexity of real domain applications.  
\end{leftbar}

There is no unified strategy for the use of existing data and the acquisition of new data for discovering functional physical matter, processes, or models.  The existence of benchmarks would encourage method development and makes it easier to evaluate when new techniques are an improvement over existing ones.  We have such evaluations for purely computational problems like numerical optimization and transition state searching, but there are no realistic benchmarks upon which to test algorithms for autonomous discovery (e.g., hypothesis generation, experimental selection, etc.).  %

\citeauthor{vempaty_coupon-collector_2017} describe one way to evaluate knowledge discovery algorithms through a simplified ``coupon-collector model''; this model assumes that domain knowledge is a set of elements to be identified through a noisy observation process \cite{vempaty_coupon-collector_2017}, which represents a limited problem formulation. 
Even for subtasks with seemingly better-defined goals like building empirical QSAR/QSPR models, %
there are no standard evaluations for assessing interpretability, uncertainty quantification, or generalizability.  
The field will need to collectively establish a set of problem formulations that describe many discovery tasks of interest to domain experts in order to benchmark components of autonomous discovery. Given the practical obstacles to validation through physical experiments, computational chemistry may be the right playground for advancing these techniques.

However, we do caution that an overemphasis on quantitative benchmarking can  be detrimental. Language tasks have reached a point where the amount of compute required for competitive performance is inaccessible for all but the most well-resourced research groups \cite{Rogers_2019_leaderboards}. Unless benchmarking controls what (open source) training data is permissible, a lack of access to compute and data may inadvertently discourage method development.

%% file: sections/conclusion.tex
\section{Conclusion}

\subsection{Attribution of discovery to automation}

The case studies in this article illustrate that computer assistance and automation have become ubiquitous parts of scientific discovery both by reducing the manual effort required to complete certain tasks and by enabling entirely new approaches to discovery at an unprecedented throughput. 
But to what extent can the discovery itself be considered a direct result of automation or autonomy? 

As summarized in our reflection of the case studies in Part 1, very few studies can claim to have achieved a high level of autonomy. In particular, researchers frequently gloss over the fact that specifying the discovery objective, defining the search space, and narrowing that space to the ``relevant'' space that is ultimately explored requires \emph{substantial} manual input. While there will always be a need for subject matter experts in constructing these platforms and associated workflows, we hope it will be possible to endow autonomous platforms with sufficiently broad background knowledge and validation capabilities that this initial narrowing of the search space is less critical to their success.

\subsection{Changing definitions of discovery}
The bar for what makes a noteworthy discovery is ever-increasing. Computer-aided structural elucidation, building structure-activity relationships, and automated reaction optimization are all discoveries under the definition we have presented here, but they are not perceived to be as significant as they were in the past. As computational algorithms become more flexible and adaptive in other contexts, and as the scope of automatable validation experiments expands, more and more workflows will appear routine.

We have intentionally avoided a precise definition of the degree of confidence required for a discovery without direct experimental observation of a desired physical property. This is because this varies widely by domain and is rapidly evolving as computational validation techniques and proxy assays become more accurate. A computational prediction of a new chemical reaction would likely not be considered a discovery under any circumstances without experimental validation. A computational prediction of a set of bioactive compounds might, but with a subjective threshold for the precision of its recommendations. Whether the computational workflow has directly made the discovery of a new compound might depend if all of the top $n$ compounds were found to be active, or if at least $m$ of the top $n$ were, etc.

\subsection{Role of the human}

The current role of humans in computer-assisted discovery is clear. Langley writes of the ``developer's role'' in terms of high-level tasks: formulating the discovery problem, settling on an effective representation, and transforming the output into results meaningful to the community \cite{langley_computer-aided_1998,langley_computational_2000}. 
\citeauthor{honavar_promise_2014} includes mapping the current state of knowledge and generating/prioritizing research questions \cite{honavar_promise_2014}.

Alan Turing's {\it imitation game} (``the Turing test'') asks whether a computer program can be made indistinguishable from a human conversationalist \cite{turing_i.computing_1950}. It is interesting to wonder if we can reach a point where autonomous platforms {\it are} able to report insights and interpretations that are indistinguishable (both in presentation and scientific merit) from what a human researcher might publish in a journal article.   Among other things, this would require substantial advances in hypothesis generation, explainability, and scientific text generation. Machine-generated review articles and textbooks may be the first to pass this test \cite{writer_lithium-ion_2019}. \citeauthor{kitano_artificial_2016}'s more ambitious grand challenge in his call-to-arms is to make a discovery in the biomedical sciences worthy of a Nobel Prize \cite{kitano_artificial_2016}.

We do not want to overstress a direct analogy of the Turing test to autonomous discoveries, because the type of discoveries typically enabled by automation and computational techniques are often distinct from those made by hand. For the field to have the broadest shared capabilities, the best discovery platforms will excel at tasks that humans can't easily or safely do.  The scale of data generation, the size of a design space that can be searched, and the ability to define new experiments that account for enormous quantities of existing information makes autonomous systems equipped to make discoveries in ways entirely distinct from humans. %

\citeauthor{turing_i.computing_1950} makes the point that the goals of machines and programs are distinct; that a human would lose in a race with an airplane does not mean we should slow down airplanes so their speeds are indistinguishable. Rephrased more recently by Steve Ley, ``while people are always more important than machines, increasingly we think that it is foolish to do thing machines can do better than ourselves'' \cite{ley_organic_2015}. Particularly when faced with the grunt work of some manual experimentation, ``leaving such things to machines frees us for still better tasks'' (Derek Lowe) \cite{lowe_derek_automated_2018}. We should \emph{embrace} the divergence of human versus machine tasks.

\subsection{Outlook}

We join many others in touting the promise of autonomous or accelerated discovery \cite{waltz_automating_2009,gil_amplify_2014,chow_streamlining_2018,alberi_2019_2018,schneider_automating_2018, saikin_closed-loop_2019,dimitrov_autonomous_2019,friederich_toward_nodate, vamathevan_applications_2019, correa-baena_accelerating_2018, tabor_accelerating_2018}. Automation has brought increased productivity to the chemical sciences through efficiency, reproducibility, reduction in error, and the ability to cope with complex problems at scale; likewise, machine learning and data science through the identification of highly nonlinear relationships, trends, and patterns in complex data. %

The previous section identified a number of directions in which additional effort is required to capture the full value of that promise: creating and maintaining high-quality open access datasets; building interpretable, data-efficient, and generalizable empirical models; expanding the scope of automated experimental platforms, particularly for multistep chemical synthesis; improving the applicability and speed of automated computational validation; aligning automated validation with prior knowledge and what is needed for different discovery applications, ideally not at significant cost; improving uncertainty quantification and cost-sensitive active learning; enabling open-ended hypothesis generation for experimental selection; and  explicitly incorporating synthesizability considerations into generative models and benchmarking on realistic tasks. Evaluation will require the creation of benchmark problems that we argue should focus on whether algorithms facilitate extrapolation to underexplored, large design spaces that are currently expensive or intractable to explore.

Numerous research initiatives are supporting work in these directions. For example, the United States Department of Defense recently funded a multidisciplinary initiative to develop a Scientific Autonomous Reasoning Agent; the Defense Advanced Research Projects Agency (DARPA) has funded several programs relevant to autonomous discovery, including the Data-Driven Discovery of Models, Big Mechanism Project, Make-It, Accelerated Molecular Discovery, and Synergistic Discovery and Design; the Engineering and Physical Sciences Research Council (EPSRC) has an ongoing Dial-a-Molecule challenge that strives to debottleneck synthesis, and recently launched a Centre of Doctoral Training in Automated Chemical Synthesis enabled by Digital Molecular Technologies; the Materials Genome Initiative, Materials Project, and Mission Innovation's Materials Acceleration Platform continue to bring sweeping changes to how data is materials science is collected, curated, and applied to discovery. Many more commercial efforts are underway as well, with significant investment from the pharmaceutical industry into the integration and digitization of their drug discovery workflows.

A 2004 perspective article by \citeauthor{glymour_automation_2004} stated that we were in the midst of a revolution to automate scientific discovery \cite{glymour_automation_2004}. Regardless of whether we were then, we certainly seem to be now.

%% file: sections/acknowledgements.tex
\section{Acknowledgements}
We thank Thomas Struble for providing comments on the manuscript and our other colleagues and collaborators for useful conversations around this topic. This work was supported by the Machine Learning for Pharmaceutical Discovery and Synthesis Consortium and the DARPA Make-It program under contract ARO W911NF-16-2-0023.